\documentclass[11pt]{article}

\usepackage{amsmath, amssymb, amsthm}
\usepackage{graphicx}
\usepackage{hyperref}
\usepackage{geometry}
\usepackage{booktabs}
\usepackage{threeparttable}
\usepackage[numbers]{natbib}
\usepackage{multirow}

\title{A New Trained Supervised Method for Calculating Patient Similarity}
\author{
  Minzee Kim\thanks{Minzee Kim is the corresponding author and may be contacted at mz2kim@uwaterloo.ca.} \\[0.5ex]
  Department of Statistics \& Actuarial Science \\ 
  University of Waterloo, Waterloo, ON, Canada, N2L 3G1\\[2ex]
  Joel A.~Dubin \\[0.5ex]
  Department of Statistics \& Actuarial Science and \\ 
  School of Public Health Sciences\\
  University of Waterloo, Waterloo, ON, Canada, N2L 3G1
}
\date{} 

\begin{document}

\maketitle

\begin{abstract}
Personalized predictive modelling has been growing rapidly with the increasing availability of Electronic Health Records. This approach aims to improve a model's predictive performance by fitting a unique model to each individual. We train the model on a subset of the training data consisting of individuals similar to the individual being predicted, identified through some similarity metric. Earlier studies show that using a personalized model trained on a customized subset of the data leads to better prediction than using a global model trained on the full dataset. In this work, we develop a new patient similarity metric to improve the prediction of a personalized model for binary response data. Specifically, we introduce a weighted cosine similarity metric that extends the standard cosine similarity metric by assigning predictor-specific weights when computing similarity between participants. These weights are estimated using a supervised approach with the relaxed adaptive group lasso. Results from simulation studies and an analysis of intensive care unit data show that although our proposed similarity metric leads to a slight deterioration in calibration, it produces substantial gains in discrimination. Overall predictive performance measured by the Brier Score improves because the increase in discrimination outweighs the loss in calibration; therefore, our proposed similarity metric more effectively identifies similar participants, resulting in improved predictive accuracy.
\end{abstract}


\maketitle

\renewcommand\thefootnote{}

\renewcommand\thefootnote{\fnsymbol{footnote}}
\setcounter{footnote}{1}

\section{Introduction} \label{sec:intro}

With the rapid development of technology, electronic health record (EHR) data has been readily available for predictive modelling. EHR data offer a huge amount of information - still, previous studies have shown that personalized predictive modelling (PPM) that uses a carefully chosen subset of the data that is more homogeneous leads to better predictive results, particularly regarding model discrimination performance, compared to using the entire data that contains heterogeneous patients \citep{ xu2019similarity, krikella2025personalized}. 

A key challenge in this area of research is quantifying the similarity between patients using their features in a given dataset to identify those in the training data that are considered to be similar to each patient in the testing data. Calculating similarity using distance-based metrics can be classified into two broad areas, where one calculates similarity using a single similarity measure, such as Euclidean distance or Mahalanobis distance, and the other uses a different feature similarity calculation for each data type and then subsequently combines the measurements \citep{wang2021study}. 

To calculate the similarity between patients, recent studies have incorporated different weights for each feature, or predictor in the EHR data, to improve the predictive performance of the model. For example, Wang et al. calculate the feature similarity for four categories - age, sex, laboratory test, and disease diagnoses - and combine them into a single metric using predetermined weights of 0.1, 0.1, 0.4, and 0.4, respectively \citep{wang2019measurement}. The weights are determined experimentally from their previous study, where they proposed three weighting schemes for the four categories  - 0.25, 0.25, 0.25, and 0.25 for the first scheme, 0.1, 0.1, 0.4, and 0.4 for the second scheme, and 0.1, 0.1, 0.6, and 0.2 for the third scheme - and the one with the best predictive performance is chosen \citep{huang2019study}. Gu et al. use an adopted weighted Euclidean distance to calculate the similarity for both continuous and discrete variables \citep{gu2017case}. The weights are derived using the genetic algorithm, which adopts the idea of natural selection and natural genetics to learn the weights based on sample cases. Gottlieb et al. categorize the features into different groups, calculate the distance between each feature with different metrics depending on the group, and combine them into a single similarity measure using a weighted geometric mean \citep{gottlieb2011predict}. Campillo-Gimenez et al. consider an exclusive OR-based patient similarity with the logistic regression model, where they use the Wald statistic of each predictor variable in the logistic regression to determine the weights of each feature \citep{campillo2013improving}. Li et al. group the features by type, such as demographics, comorbidities, lab tests, and procedures and use predefined empirical weights, expert-assigned weights, and different grid-search weight combinations before combining them into a single scalar similarity metric \citep{li2025comparative}. Lambert et al. use a weighted cosine similarity metric to build a patient network using the Markov Cluster Algorithm, where they assign different weights to the features \citep{lambert2025improving}. The cosine similarity is weighted by Inverse Document Frequency (IDF), Wu \& Palmer measure, or by the Lin measure \citep{sparck1972statistical, wu1994verb, lin1998information}. However, these weights are not learned or optimized but are calculated mathematically from the hierarchical structure of the feature and their frequency. Some papers approach calculating patient similarity using feature selections, where some selected features have a weight of 1 and other unselected features have a weight of 0. For example, Suo et al. calculate the similarity using the patients' predicted future feature values and apply group lasso to determine these features \citep{Suo2018Multi-task}. Pai and Bader use machine learning algorithms that use a network of features to tune for the optimal selection of the features that have the highest discriminative power \citep{Pai2018Patient}.   

We hope to contribute to this area of statistics by developing a new method to calculate patient similarity. In particular, we introduce a new supervised measure of patient similarity that uses a regularization technique to develop a new weighted cosine similarity metric. Cosine similarity metric (CSM) is an effective unsupervised measure of similarity, expressed as 
\begin{equation} \label{csm}
CSM(S_i, S_j) = \frac{\sum^P_{p=1}  S_{ip} \times S_{jp}}{\sqrt{ \sum^P_{p=1} S_{ip}^2} \sqrt{\sum^P_{p=1} S_{jp}^2 }},
\end{equation}
where $P$ is the total number of features in the vector for each participant $\mathbf{S}_i$ and $\mathbf{S}_j$ \citep{li2013distance}. However, it can be overly biased by features of higher values and does not take the number of shared features between the two vectors into account \citep{li2013distance}. For example, consider the following three vectors, $A = (1, 2, 0)^T, B = (0, 1, 0)^T$, and $C = (1, 2, 2)^T$. When we consider the similarity between $A$ and $B$ and $A$ and $C$, we might say that $C$ is more similar to $A$ than $B$ since the first two features of $A$ and $C$ are equal. However, $CSM(A, B) = 0.894$ and $CSM(A, C) = 0.745$, and so $A$ and $B$ are more similar in terms of the cosine similarity metric \citep{li2013distance}. 

Our method in this paper adapts the soft similarity measure developed by Sidorov et al. and updates the existing cosine similarity metric with weights derived from regularization, further described in Section~2.1 \citep{sidorov2014soft}. Unlike existing weighted cosine similarity methods, our proposed weighted CSM assigns supervised weights and does not involve categorizing different types of predictors into groups to assign weights depending on the group characteristic. Thus, we do not require the subject matter expert's opinion to create clusters of different features. Instead, the weights are derived from the predictor's relationship with the outcome of interest by fitting a model with all potential predictors and then identifying those that are correlated with the outcome. Furthermore, our method effectively assigns weights of 0 to less significant features and assigns flexible weights to the predictors that are selected for the model. In Section~2, we introduce the algorithm to calculate the weighted CSM with details on the regularization, along with the algorithm to implement PPM with the new similarity measurement. In Sections~3 and~4, respectively, we illustrate the improved predictive performance of a statistical model with our proposed method through a simulation study and data analysis using the publicly-available eICU Collaborative Research Database \citep{pollard2018eicu}. We conclude with a discussion and some suggestions on future work in Section~5.

\section{Methods}\label{sec:methods}

\subsection{Weighted Cosine Similarity Metric} \label{sec:Weighted cosine similarity metric}
We propose a new weighted cosine similarity metric to quantify the similarity between two models in a vector space model, which can be implemented in personalized predictive modelling (PPM) as well to measure the similarity between two individuals \citep{sidorov2014soft}:

\begin{equation} \label{weighted csm}
CSM_w(S_i, S_j) = \frac{\sum^P_{p=1} w_p \times S_{ip} \times S_{jp}}{\sqrt{ \sum^P_{p=1} w_p \times S_{ip}^2} \sqrt{\sum^P_{p=1} w_p \times S_{jp}^2 }}.
\end{equation}

We introduce two new approaches to determine $w_p$ that allow for curvilinear or non-linear representations of some variables in the linear predictor. The regularization technique that we use to compute $w_p$ is the relaxed adaptive group lasso, which enables determining $w_p$ supervised without requiring further knowledge about the types of predictors or expert knowledge, as previously published \citep{wang2019measurement}. Note that while relaxed adaptive group lasso is not a new contribution to the literature, to our knowledge, it has not been used before to compute the weights of the weighted CSM to be used in the context of PPM.

Since we have a large number of predictors to calculate the similarity between the participants in the data, it is common to consider a regularization method to retain only a subset of the predictors. We consider the relaxed adaptive group lasso for several reasons. First, lasso (least absolute shrinkage and selection operator) is a generally effective regularization approach to reduce the dimension of the parameter space in modelling (often high-dimensional) data, particularly using generalized linear models \citep{hastie2009elements}. However, the model selection result could be inconsistent, mostly because lasso applies the same amount of shrinkage to each coefficient. As a solution, adaptive lasso is introduced to allow for different amounts of shrinkage on different regression coefficients, thereby reducing bias \citep{zou2006adaptive}. Adaptive group lasso is a natural extension of adaptive lasso, best suited for situations where it is desirable to select the predictors in a grouped manner instead of selecting them individually. For example, a collection of indicators representing a categorical predictor with three levels should be included or excluded from the model as a whole, rather than having only some of its indicator variables selected. It allows for different tuning parameters to be used for different groups of predictors to shrink a subset of coefficients to zero \citep{wang2008note}. Mathematically, let $y \in \mathbb{R}^n$ denote the outcome, or response, vector and 
$X \in \mathbb{R}^{n \times p}$ the design matrix.
The $p$ predictors are partitioned into $m$ non-overlapping groups
$\{G_1, \dots, G_m\}$, where $G_g \subset \{1,\dots,p\}$ and $|G_g| = p_g$.
Let $\beta = (\beta_{G_1}^\top, \dots, \beta_{G_m}^\top)^\top \in \mathbb{R}^p$
denote the corresponding coefficient vector. $\beta$ is estimated using the adaptive group lasso as 
\begin{equation}
\hat{\beta} =
\arg\min_{\beta \in \mathbb{R}^p}
\left\{
\frac{1}{2}\|y - X\beta\|_2^2
+
\lambda
\sum_{g=1}^m
w_g \, \|\beta_{G_g}\|_2
\right\},
\label{eq:agl}
\end{equation}
where the adaptive weights are defined as
\begin{equation}
w_g
=
\left(
\|\tilde{\beta}_{G_g}\|_2 
\right)^{-\gamma},
\quad \gamma > 0,
\label{eq:weights}
\end{equation}
where $\tilde{\beta}$ is a group lasso estimator \citep{wang2008note}.

Finally, we extend this to relaxed adaptive group lasso, which can reduce estimation bias while maintaining sparsity by refitting a model on the predictors selected by the adaptive group lasso \citep{meinshausen2007relaxed}.

Let the set of selected groups from adaptive group lasso be defined as
\begin{equation}
\mathcal{A}
=
\left\{
g : \|\hat{\beta}_{G_g}\|_2 > 0
\right\}.
\label{eq:active}
\end{equation}

Conditioning on the selected group set $\mathcal{A}$, we obtain the relaxed adaptive group lasso estimator by solving
\begin{equation}
\hat{\beta}^{\mathrm{RAGL}}
=
\arg\min_{\beta \in \mathbb{R}^p}
\left\{
\frac{1}{2}\|y - X\beta\|_2^2
+
\lambda_1
\sum_{g \in \mathcal{A}}
w_g \, \|\beta_{G_g}\|_2
\right\},
\label{eq:ragl}
\end{equation}
subject to $\beta_{G_g} = 0 \;\; \forall g \notin \mathcal{A}$ where $0 \le \lambda_1 \le \lambda$.
A common special case is $\lambda_1 = 0$, corresponding to refitting the model using only the predictors in $\mathcal{A}$. Applying the relaxed adaptive group lasso strategy, we introduce two new weighted CSM methods as follows:

\begin{enumerate}
    \item Created an augmented design matrix $\mathbf{X^*}$ that contains the original values for the predictors in the data and non-linear transformations of the predictors. For example, this could be an augmented design matrix of the original predictors, squared values of the predictors, and/or the exponential of the values of the predictors. This is to account for real-life scenarios where we do not know the true outcome model, and some variables have curvilinear or non-linear representations in the linear predictor. An example of this is demonstrated in the simulation study in Section~3.

    \item Standardize each column of $\mathbf{X^*}$.

    \item Define the grouping structure for the augmented design matrix $\mathbf{X}^*$. Let $P$ denote the number of original predictors. For each original predictor $p \in \{1, \dots, P\}$, define a group $G_p \subset \{1, \dots, p^*\}$ that contains the column index corresponding to predictor $p$ and the column indices of all its associated nonlinear transformations in $\mathbf{X}^*$, where $p^*$ denotes the total number of columns in $\mathbf{X}^*$. Thus, the augmented predictors are partitioned into $G = P$ non-overlapping groups $\{G_1, \dots, G_P\}$, with each group corresponding to one original predictor and its transformations.

    \item Perform relaxed adaptive group lasso as described in Section~2.1.

    \item Using the estimated coefficients from the relaxed adaptive group lasso, we define predictor-specific weights $w_p$ for $p = 1, \dots, P$ to construct the weighted cosine similarity metric. Let the set of selected predictor groups at tuning parameter value $\lambda$ be defined as
    \begin{equation}
    \mathcal{G}_\lambda=\left\{p \in \{1,\dots,P\} :\|\hat{\beta}_{G_p,\lambda}\|_2 > 0\right\}.
    \end{equation}
    \begin{enumerate}
    \item \textbf{Method 1 (weight01 CSM).} Define
    \begin{equation}
    w_p =
    \begin{cases}
    1, & p \in \mathcal{G}_\lambda, \\
    0, & p \notin \mathcal{G}_\lambda.
    \end{cases}
    \end{equation}
    That is, all predictors whose corresponding groups are selected by the adaptive group lasso are weighted equally, while unselected predictors receive zero weight.

    \item \textbf{Method 2 (weightbeta CSM).} For each selected predictor group $p \in \mathcal{G}_\lambda$, define a group-level importance score
    \begin{equation}
    s_p=\|\hat{\beta}_{G_p,\lambda}\|_2=
    \left(
    \sum_{q \in G_p}
    \hat{\beta}_{q,\lambda}^2
    \right)^{1/2}.
    \end{equation}
    The predictor-specific weights are then given by
    \begin{equation}
    w_p =
    \begin{cases}
    0, & p \notin \mathcal{G}_\lambda, \\[6pt]
    \displaystyle
    \frac{s_p}{\sum_{p' \in \mathcal{G}_\lambda} s_{p'}},
    & p \in \mathcal{G}_\lambda.
    \end{cases}
    \end{equation}
    \end{enumerate}

\end{enumerate}

For convenience, especially in the simulation section, we refer to Method 1 as ``weight01 CSM", since it is a weighted CSM method with a weight of 1 for the selected predictors and a weight of 0 for the non-selected predictors, and Method 2 as ``weightbeta CSM" since it is a weighted CSM method where the nonzero weights are calculated using the beta estimates. To compare the two new proposed methods, we refer to ``standard CSM" as the regular CSM introduced in the Introduction that uses all available predictors with an equal weight of 1 across all predictors.

\subsection{Algorithm for PPM}\label{sec:Algorithm for ppm}

In this section, we describe our new proposed algorithm to improve the predictive performance of generalized linear models with the weighted CSM calculation described in Section~2.1.

\begin{enumerate}
    \item Split the data into training and testing set (TrTe data) and validation set (Validation data), where $100 \times q$\% of the data is used as TrTe data and the remaining $100 \times (1-q)$\% of the data is used as the Validation data. 

    \item Split the TrTe data into $K$ folds for $K$-fold cross-validation (CV).
    
    \item Let $k$ be the hold-out fold for testing data, and the other $K-1$ folds be the training data.

    \item Consider a grid of $M_p$ values to tune $M_p$.  

    \begin{enumerate}
        \item For an index participant $j$ in the testing data, apply the weighted CSM algorithm described in Section~2.1 to the training dataset.
        \item Sort the participants in the training dataset in descending order by the similarity metric.
        \item Fit a generalized linear model on the top $M_p \times 100$\% of the sorted training data and estimate the outcome for participant $j$. 
        \item Repeat Steps 4(a) - 4(c) for each participant $j$ in the testing data and calculate the model performance measures of interest, such as discrimination, calibration, or Brier score. 
    \end{enumerate}

    \item Repeat Steps 3-4 for each $k$ in $K$ folds. Average over the values of performance metrics over the $K$ folds.

    \item Repeat Steps 2-5 for all $K$ folds and repeat the entire $K$-fold cross-validation procedure $R$ times. This leads to a total of $R \times K$ evaluations, each using a different training-testing split. The performance metric is averaged over all $R \times K$ evaluations, and the optimal $M_p$ is selected as the value achieving the best averaged performance according to a pre-specified evaluation metric. In our simulation study and data analysis, we use Brier score as the evaluation metric since it evaluates overall predictive accuracy, incorporating both discrimination and calibration \citep{murphy1973new}.

    \item We validate the model using the Validation data obtained in Step 1.

    \begin{enumerate}
        \item From the Validation data, create a bootstrap sample of the same length by sampling the data rows with replacement. 
        \item Randomly split the bootstrap sample into $80$\% training sample and $20$\% testing sample.
        \item For each participant $j$ in the testing sample, measure the similarity between participant $j$ and all the participants in the training sample.
        \item Sort the participants in the training sample by descending order of similarity. Using the winning $M_p$ value obtained from Steps 2-6, obtain a subpopulation of the training set that is the most similar to participant $j$ from the testing set. 
        \item Evaluate the predictive performance by fitting a generalized linear model on the subpopulation obtained in Step 7(c), for each $j$ in the testing sample.
        \item Repeat Steps 7(a)-7(e) B number of times to obtain a bias-corrected and accelerated (BCa) bootstrap confidence interval around the performance measure of the validation sample. 
    \end{enumerate}

\end{enumerate}

\section{Simulation Studies}\label{sec:simulations}

In this section, we describe a simulation study conducted on various datasets to investigate if using our proposed weighted CSM can improve the predictive performance of models in PPM. We follow the proposed algorithm outlined in Section~2.2 to implement a 5-fold cross-validation repeated twenty times to tune for the best subpopulation proportion for the dataset generated in this section. We consider the following subpopulation proportion sizes $M_p$ - $0.1, 0.2, 0.3, 0.4, 0.5, 0.6, 0.7, 0.8, 0.9$, and $1$ - and use both discrimination and calibration to evaluate the predictive performance of the model. Note that the subpopulation proportion of $1$ is equivalent to using all available data in the training data to fit the same global model for all participants in the testing data. In the validation step, we use $B = 500$ bootstrap samples to generate the bootstrap confidence interval around the performance measures of the validation sample. Our algorithm is tested on five different datasets generated.

Area Under the Receiver Operating Characteristic Curve (AUROC) and Area under the Precision-Recall Curve (AUPRC) are considered to assess discrimination. Receiver Operating Characteristic Curve displays the trade-off between sensitivity and specificity across all possible threshold values, and AUROC is the area under this curve that ranges between 0.5 and 1 \citep{ozenne2015precision}. AUPRC is an alternative approach to measuring discrimination that uses the precision-recall curve that shows the trade-off between precision and sensitivity (also called recall), where precision is the number of true positives divided by the total number of predicted positives. Unlike AUROC, possible values of AUPRC depend on the prevalence of the data and decrease towards 0 as the prevalence gets lower \citep{ozenne2015precision}. To evaluate discrimination, we calculate AUPRC for datasets with lower prevalence and AUROC for datasets with higher prevalence because AUPRC is a more informative measure of the performance of the model with highly imbalanced data \citep{hancock2023evaluating}. For both measures, high values indicate better predictive performance. We use the Integrated Calibration Index (ICI) to assess calibration, which is a measure of moderate calibration that represents the weighted average absolute difference between the observed and predicted probabilities \citep{austin2019integrated}. That is, it measures the weighted difference between observed and predicted probabilities, where the observations are weighted by the empirical density function of the predicted probabilities. Finally, to consider a combination of both discrimination and calibration, we use the Brier score (BrS) which is an overall predictive performance measure that can be decomposed into both discrimination and calibration \citep{murphy1973new}.

We begin with Section~3.1 where we describe in detail how the different datasets are generated. In Section~3.2, we discuss the simulation results comparing our proposed methods, weight01 CSM and weightbeta CSM, with standard CSM.

\subsection{Setup} \label{sec:Setup}
We consider six cases for this simulation study, coded in R version 4.4.0 \citep{R} using RStudio \citep{Rstudio}. In all six cases, we generate the data using a sample size of $n = 10000$ with 20 predictors. 15 of these predictors are continuous variables generated using an independent normal distribution of $N(0, 2^2)$ and five are binary predictors with an independent Bernoulli distribution of $Bernoulli(0.5)$. Among these 20 predictors, seven continuous predictors and three binary predictors are used to generate a binary outcome $Y$ for each sample in the data with $\varepsilon \sim N(0, 2.5^2)$. We consider 0.1 and 0.5 as options for the prevalence of the outcome of the data set, and three different settings (weak signal, moderate signal, and strong signal) for the strength of the association between the predictors and the outcome. Specifically, the linear predictors for the six cases are as follows:

\textbf{Case 1: Prevalence = 0.1, weak signal}
\begin{align}
    z &= -2 + 1x_1 + 0.5x_2 -0.5x_3 -1x_4 + 0.5x_5 \notag \\ 
      &- 0.5 \exp(x_6) - 1x_7^2 -1x_{16} - 1x_{17} + 0.5x_{18} + \varepsilon
\end{align}

\textbf{Case 2: Prevalence = 0.1, moderate signal}
\begin{align}
    z &= -3 + 3x_1 + 1.5x_2 -1.5x_3 -3x_4 + 1.5x_5 \notag \\ 
    &- 1.5 \exp(x_6) - 3x_7^2 -3x_{16} - 3x_{17} + 1.5x_{18} + \varepsilon
\end{align}

\textbf{Case 3: Prevalence = 0.1, strong signal}
\begin{align}
    z &= -4.5 + 5x_1 + 3x_2 -3x_3 -5x_4 + 3x_5 \notag \\ 
    &- 3 \exp(x_6) - 5x_7^2 -5x_{16} - 5x_{17} + 3x_{18} + \varepsilon
\end{align}

\textbf{Case 4: Prevalence = 0.5, weak signal}
\begin{align}
    z &= -0.75 + 1x_1 + 0.5x_2 +0.5x_3 +1x_4 + 0.5x_5 \notag \\ 
    &- 0.5 \exp(x_6) + 1x_7^2 -1x_{16} - 1x_{17} + 0.5x_{18} + \varepsilon
\end{align}

\textbf{Case 5: Prevalence = 0.5, moderate signal}
\begin{align}
    z &= -1.75 + 3x_1 + 1.5x_2 +1.5x_3 +3x_4 + 1.5x_5 \notag \\ 
    &- 1.5 \exp(x_6) + 3x_7^2 -3x_{16} - 3x_{17} + 1.5x_{18} + \varepsilon
\end{align}

\textbf{Case 6: Prevalence = 0.5, strong signal}
\begin{align}
    z &= -2.5 + 5x_1 + 3x_2 +3x_3 +5x_4 + 3x_5 \notag \\ 
    &- 3 \exp(x_6) + 5x_7^2 -5x_{16} - 5x_{17} + 3x_{18} + \varepsilon
\end{align}

To generate the binary outcome using the $z$ above from the six cases, we use
\begin{equation}
    \pi = 1/(1+\exp(-z))
\end{equation}
\begin{equation}
    Y \sim Bernoulli(\pi).
\end{equation}

We fit a logistic regression model using all 20 predictors, no interactions, and no nonlinear transformations of the predictors to predict the outcome of the index patient, thus introducing multiple forms of misspecification at the model fitting stage (i.e., too many noise predictors and ignoring nonlinearities for two of the signal predictors).

\subsection{Simulation Results} \label{sec:Simulation Results}
In this section, we present the results of the simulation study. To investigate the effectiveness of weight01 CSM and weightbeta CSM in improving the predictive performance of PPM, we compare our proposed similarity metrics with the following existing similarity calculations: Euclidean distance \citep{fang2021patient}, Mahalanobis distance \citep{mahalanobis2018generalized}, standard CSM, and weighted CSM with the weights determined using the Wald statistic (referred to as weightwald CSM onward). More specifically, these weights are determined using the Wald statistic of each predictor variable after fitting a logistic regression model on the available training dataset \citep{campillo2013improving}. Note that unlike our proposed weight01 CSM and weightbeta CSM methods, weightwald CSM uses all of the available predictor variables in the data without incorporating feature selection.  

Although the algorithm was evaluated on five independently generated datasets for each of the six cases, we focus on one dataset per case, since the results were consistent across all datasets. For the six cases, we display plots showing AUROC or AUPRC, ICI, and BrS across ten different subpopulation proportions to tune for the optimal subpopulation proportion using the TrTe data as described in Section~2.2. Note that for AUROC and AUPRC, a greater value represents better model discrimination; for ICI, a lower value represents better model calibration, and for Brier score, a lower score represents a better summary measure of model prediction. We implement a 5-fold CV repeated twenty times, meaning that we can also find the standard error and obtain an approximate $100 \times (1-\alpha)\%$ confidence interval around the estimate of the performance measures for each subpopulation proportion using the following equation \citep{bates2023cross, hastie2009elements}:
\begin{equation} \label{CI}
    \left (\bar{e} - z_{1 - \alpha/2} \widehat{SE}, \bar{e} + z_{1 - \alpha/2} \widehat{SE} \right),
\end{equation}
where 
\begin{equation} \label{SE}
    \widehat{SE} = \frac{1}{\sqrt{n}} \times \sqrt{\frac{1}{n-1} \sum^n_{i=1}(e_i - \bar{e})^2}
\end{equation}
such that $n$ is the number of folds (100 in this scenario), $e_i$ is the estimate from the $i$th fold, $\bar{e}$ is the average of $e_i$ across all folds, and $z_{1 - \alpha/2}$ is the $ 1 - \alpha/2$ quantile of the standard normal distribution. We will mention an improvement we can make to this approximate interval in Section~5. Note that the general shape of the plots for Cases 1, 2, and 3 is similar to each other, and the plots for Cases 4, 5, and 6 are similar to each other. Therefore, we discuss Case 2 (moderate signal with low prevalence) and Case 5 (moderate signal with higher prevalence) in more detail below and present the plots for Cases 1, 3, 4, and 6 in the Appendix Section A. Next, we display the results from the validation study using the optimal subpopulation proportion found in the tuning process for the four different metrics for Case 2 and Case 5. The interval around each estimate is found using the bias-corrected and accelerated bootstrap intervals. Finally, we present two tables showing the relative difference in the performance metrics for weightbeta CSM and weight01 CSM relative to the standard CSM for all six cases.

\begin{figure}[htbp]
\centering

\begin{minipage}[t]{0.49\textwidth}
    \centering
    \includegraphics[width=\textwidth]{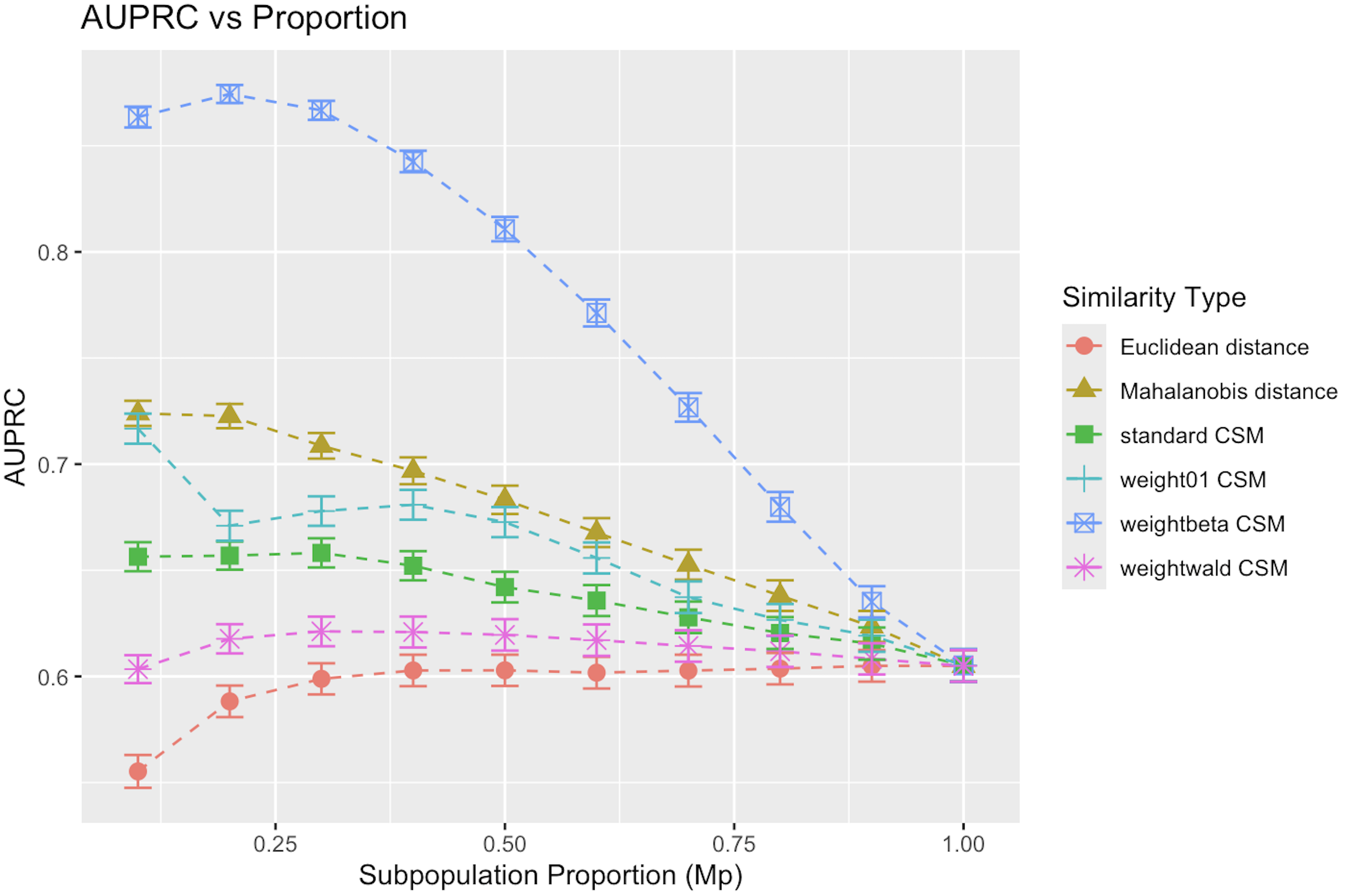}
    \label{fig:plot1}
\end{minipage}
\hfill
\begin{minipage}[t]{0.49\textwidth}
    \centering
    \includegraphics[width=\textwidth]{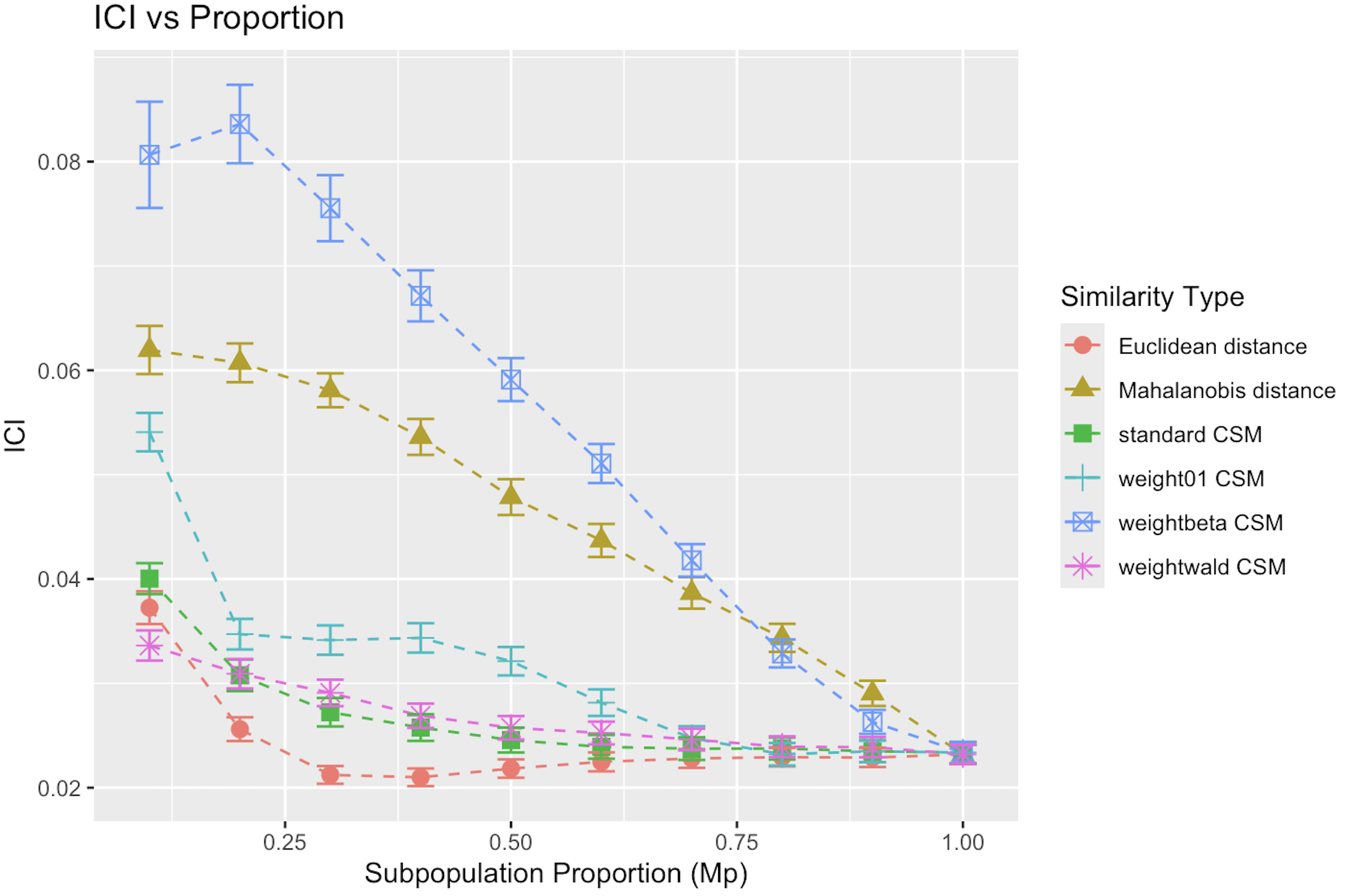}
    \label{fig:plot2}
\end{minipage}

\vspace{0.5cm}

\begin{minipage}[t]{0.49\textwidth}
    \centering
    \includegraphics[width=\linewidth]{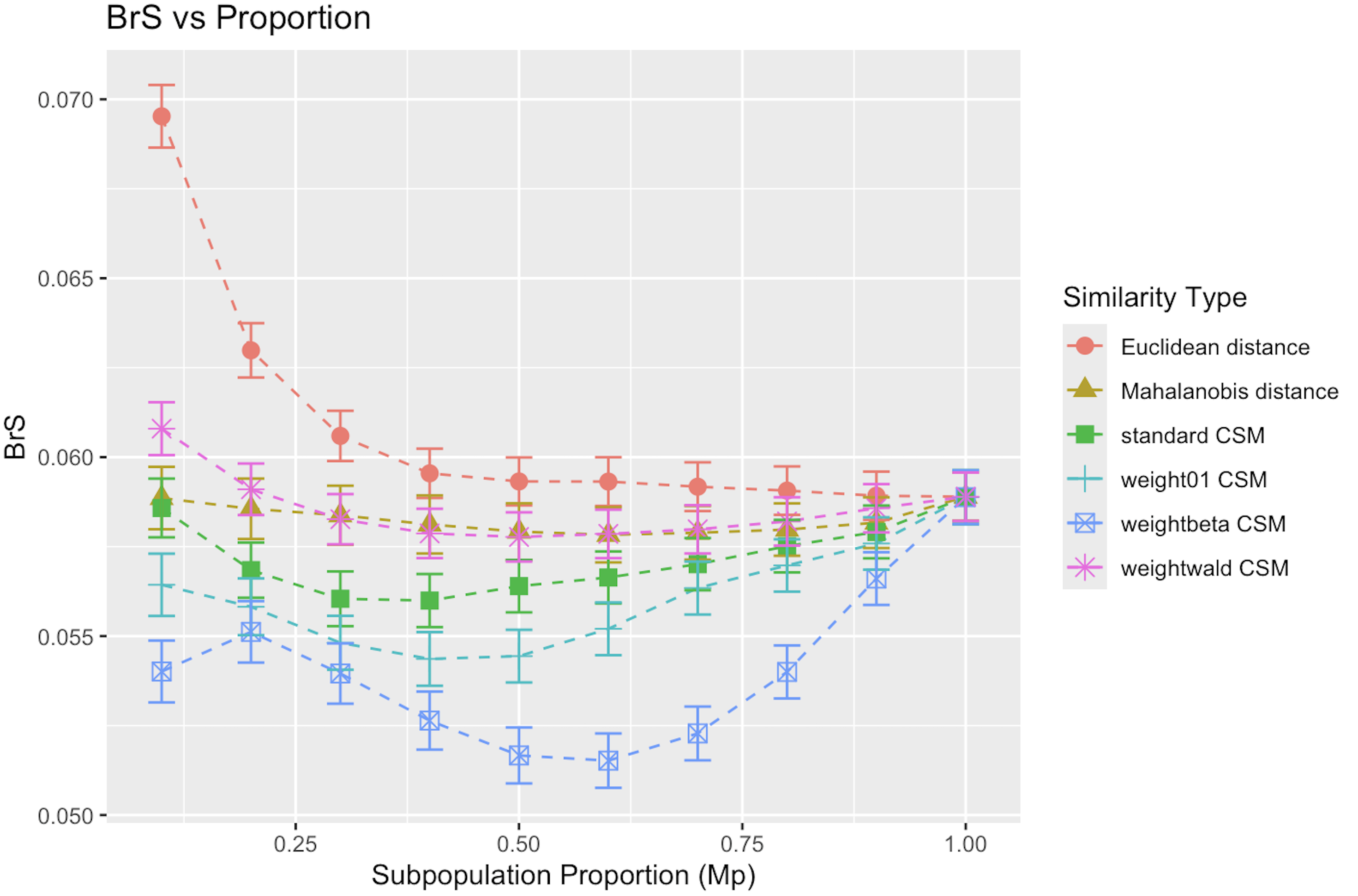}
\end{minipage}\hfill%
\begin{minipage}[t]{0.5\textwidth}
\centering
    \includegraphics[width=\textwidth]{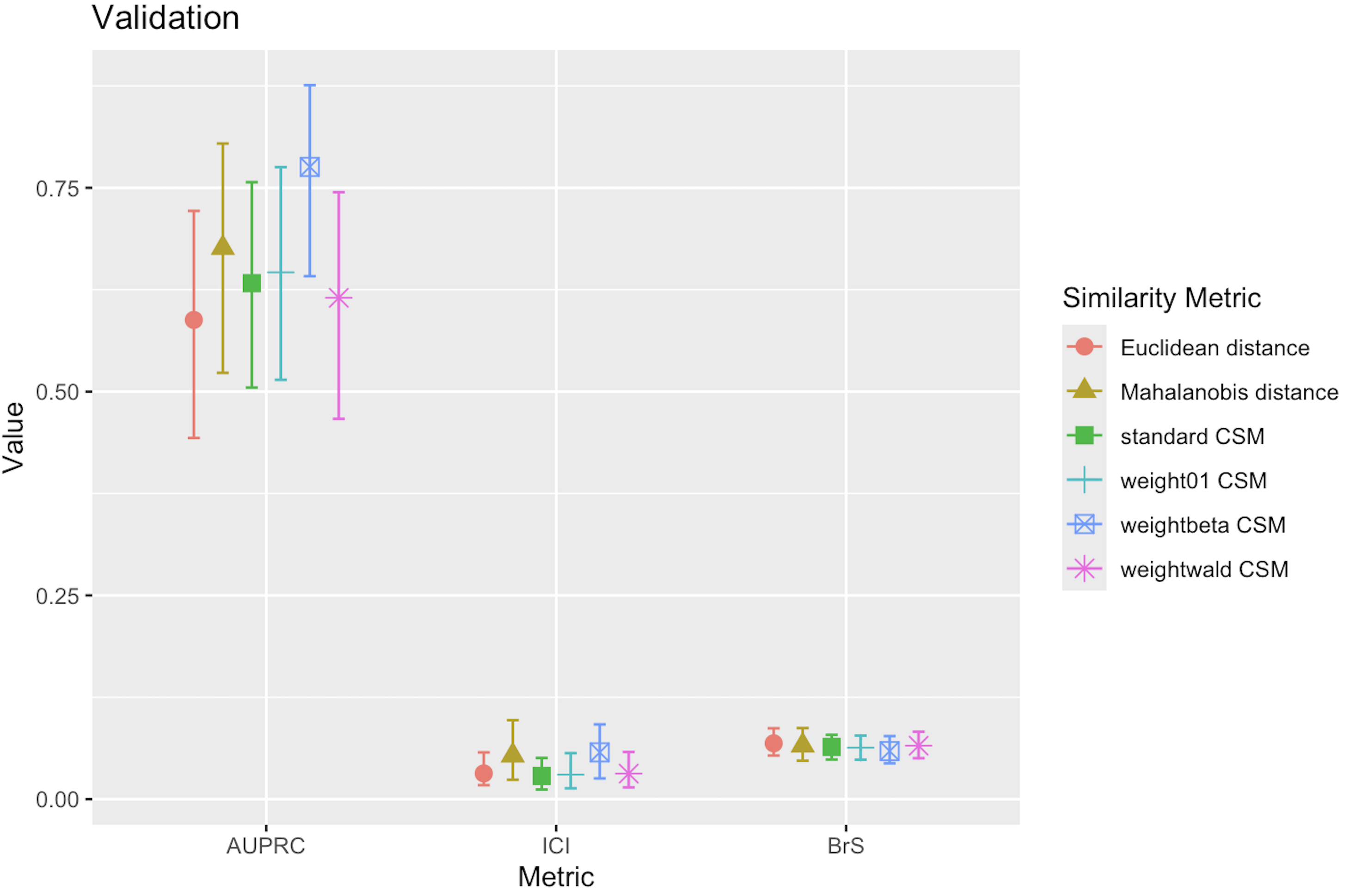}
    \label{fig:plot2}
\end{minipage}

\caption{Performance metrics across different subpopulation proportions for the six similarity measures. The dataset has low prevalence with moderate signal (Case 2).}
\label{fig:4plotscase2}
\end{figure}

\begin{figure}[htbp]
\centering

\begin{minipage}[t]{0.49\textwidth}
    \centering
    \includegraphics[width=\textwidth]{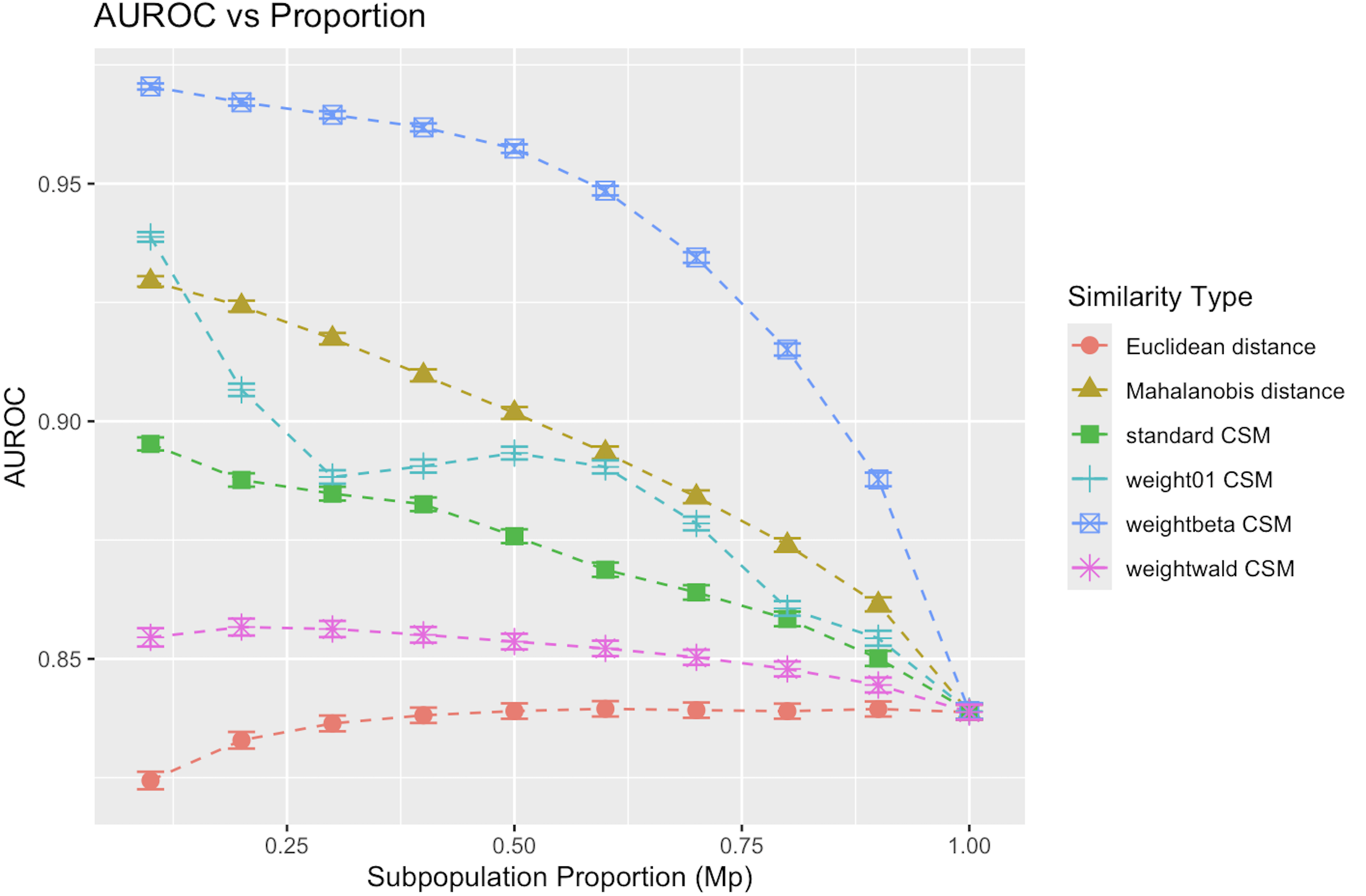}
    \label{fig:plot1}
\end{minipage}
\hfill
\begin{minipage}[t]{0.49\textwidth}
    \centering
    \includegraphics[width=\textwidth]{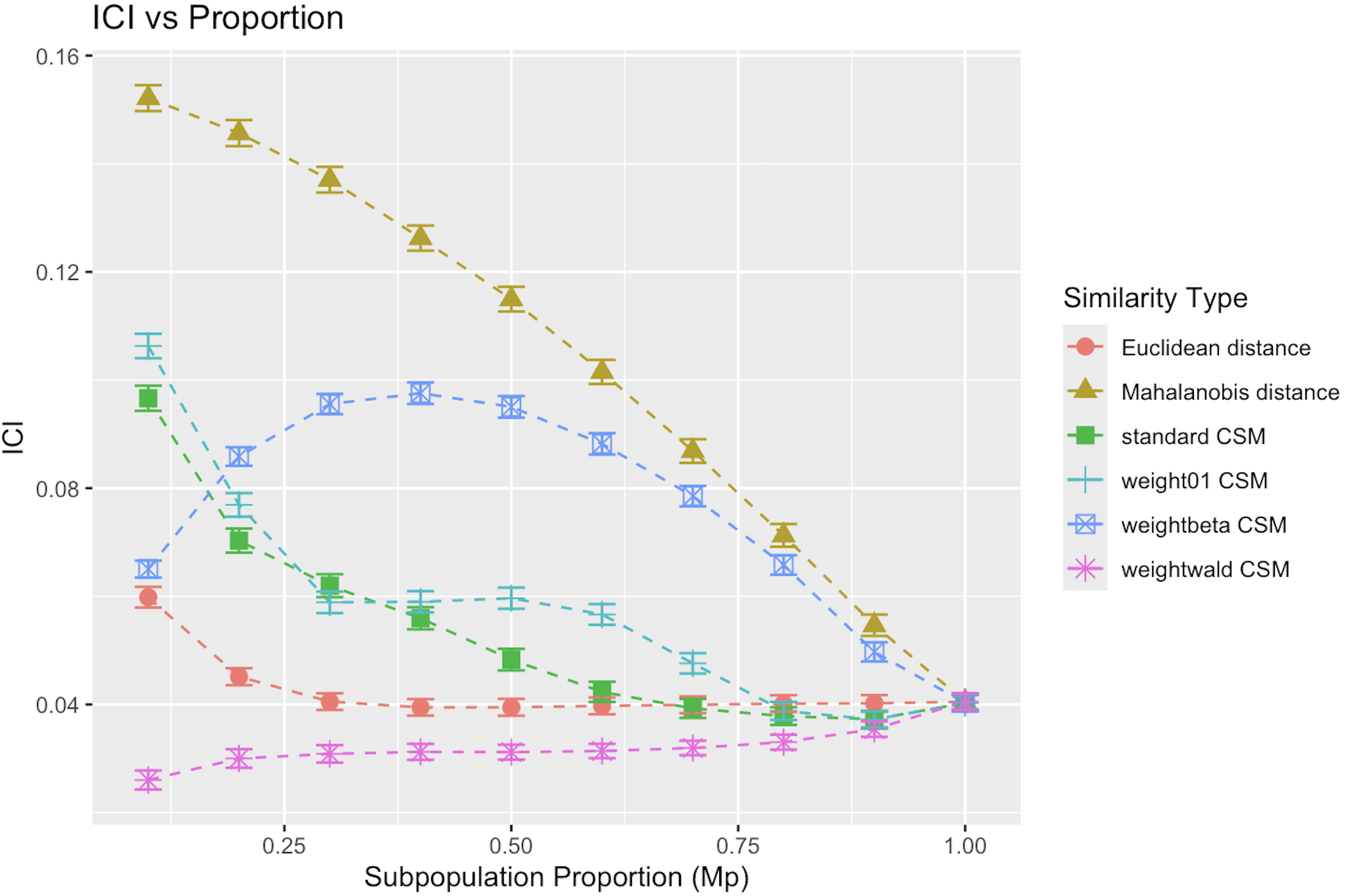}
    \label{fig:plot2}
\end{minipage}

\vspace{0.5cm}

\begin{minipage}[t]{0.49\textwidth}
    \centering
    \includegraphics[width=\linewidth]{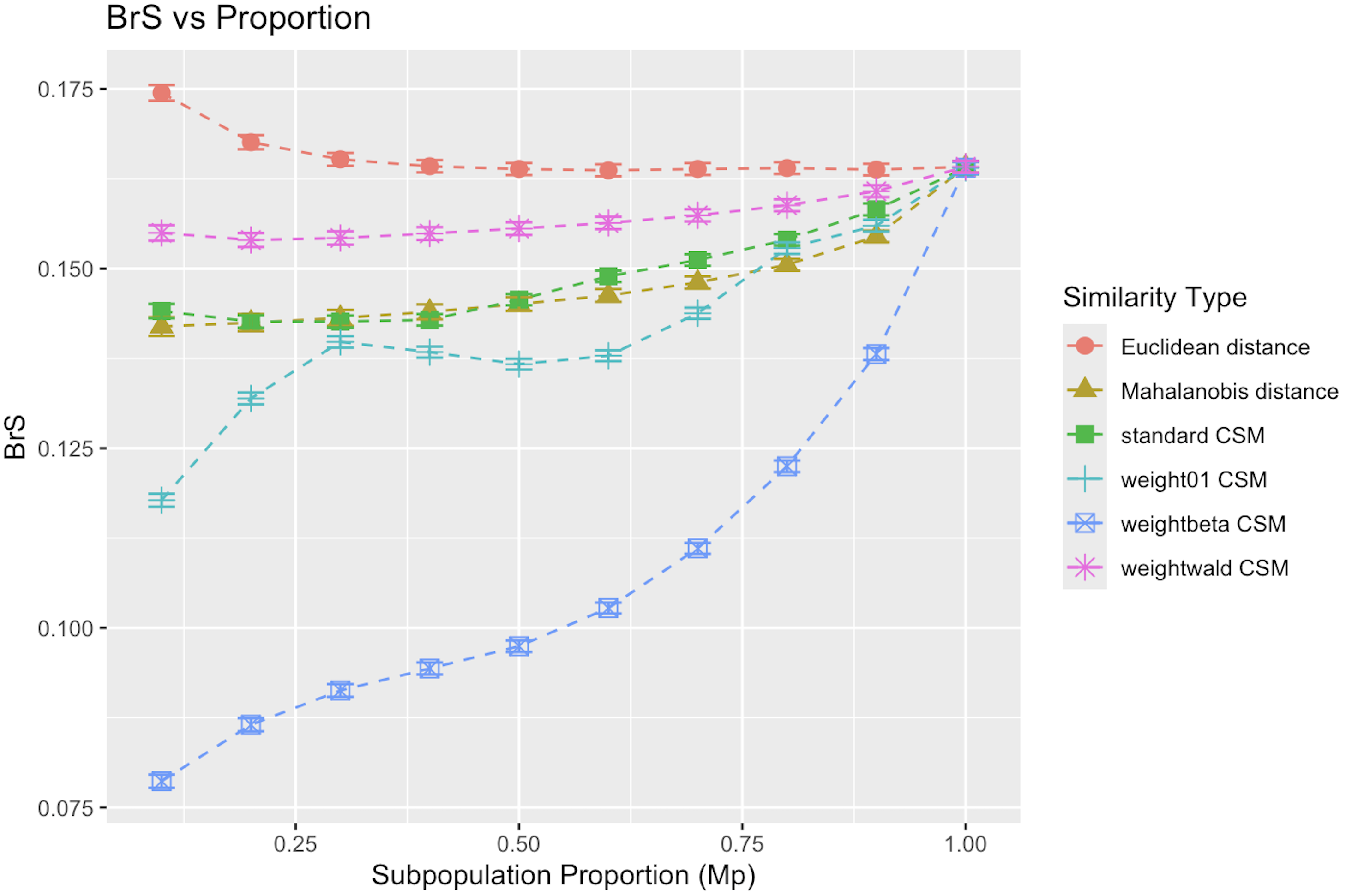}
\end{minipage}\hfill%
\begin{minipage}[t]{0.5\textwidth}
\centering
    \includegraphics[width=\textwidth]{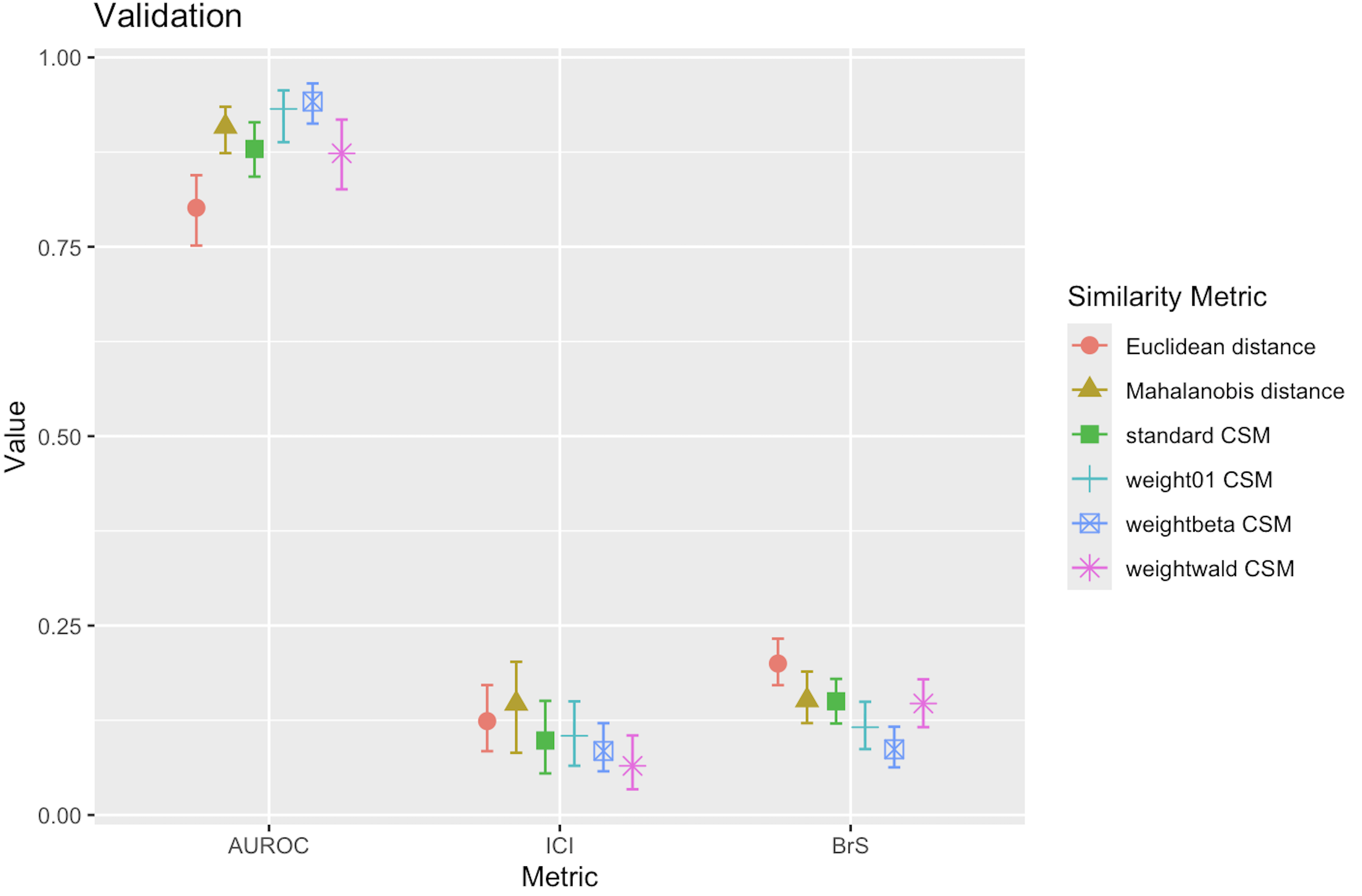}
    \label{fig:plot2}
\end{minipage}

\caption{Performance metrics across different subpopulation proportions for the six similarity measures. The dataset has a reasonably high prevalence with moderate signal (Case 5).}
\label{fig:4plotscase5}
\end{figure}

The simulation results of Case 2, shown in Figure~1, agree with the existing literature that we can improve discrimination while training the model on targeted and more homogeneous data, specific to each person in the testing data. The first plot comparing AUPRC across different subpopulation proportions shows that as the subpopulation proportion increases, AUPRC decreases. This effect is amplified with our proposed methods - when we assign a weight of 0 or 1 on the features to calculate similarity (weight01 CSM), the model consistently has better discrimination than standard CSM, weightwald CSM, and Euclidean distance. When we assign weights derived from the beta estimates from relaxed adaptive group lasso to calculate the similarity (weightbeta CSM), we observe a large improvement in model discrimination for all subpopulation proportions compared to all other similarity metrics in this simulation study. In particular, we can achieve maximum discrimination when we use the top 20\% of the most similar data to train the model to predict the outcome. However, with calibration measured by ICI, different patterns are observed. First, we see that generally, training the data on a targeted subpopulation does not have a large improvement compared to fitting a global model among standard CSM, weightwald CSM, weight01 CSM, and weightbeta CSM. Among the compared methods, Euclidean distance shows the biggest improvement in calibration when using a targeted subset of the data to fit the prediction model compared to using a global model. However, there is still a very slight improvement, especially with using weight01 CSM to calculate the similarity - training the model on a subpopulation proportion of 80\% has a negligibly slightly better calibration at 0.0232 than using the global dataset with a calibration of 0.0234. Another pattern to note is that standard CSM generally performs better than weight01 CSM and weightbeta CSM across all subpopulation proportions except at 80\%, where weight01 CSM outperforms standard CSM. For BrS, which incorporates both discrimination and calibration, we see that the best result is achieved with weightbeta CSM when the subpopulation proportion is 60\%. Weightbeta CSM consistently outperforms weight01 CSM, and weight01 CSM consistently outperforms all other methods, including Euclidean distance, Mahalanobis distance, standard CSM, and weightwald CSM. This result is as expected - BrS can be decomposed into discrimination, where weightbeta CSM greatly improves discrimination compared to other similarity measures for smaller subpopulation proportions, and calibration, where the model had better predictive performance with greater subpopulation proportion; the large gains in discrimination due to the weightbeta approach appear to outweigh the less impressive performance in calibration, leading to a noticeable improvement in Brier score.

After tuning for the optimal subpopulation proportion, which is 0.6 determined by BrS, we validate the model performance using the holdout validation set and compare AUPRC, ICI, and BrS using the six similarity metrics. These results are shown in Table~1, which includes the estimate of the performance metrics along with the bias-corrected and accelerated bootstrap intervals. We can see that when we use a subpopulation proportion of 0.6, weightbeta CSM outperforms all other methods in AURPC and BrS. 

\begin{table}[h] 
\caption{Results from the validation study for low prevalence, moderate signal (Case 2). The columns ``Lower'' and ``Upper'' represent the 2.5\% BCa and 97.5\% BCa bootstrap interval estimates, respectively, and the winning $M_p$ from the tuning process used to calculate the performance metrics is $0.6$.}
\centering
\begin{tabular}{|l|l|l|l|l|}
\hline
\multicolumn{1}{|c|}{Metric} & \multicolumn{1}{c|}{Similarity metric} & \multicolumn{1}{c|}{Est} & \multicolumn{1}{c|}{Lower} & \multicolumn{1}{c|}{Upper} \\ \hline

\multirow{6}{*}{AUPRC}
 & Euclidean distance        
 & \begin{tabular}[c]{@{}l@{}}0.588\end{tabular}
 & \begin{tabular}[c]{@{}l@{}}0.443\end{tabular}
 & \begin{tabular}[c]{@{}l@{}}0.722\end{tabular} \\
 
 & Mahalanobis distance        
 & \begin{tabular}[c]{@{}l@{}}0.676\end{tabular}
 & \begin{tabular}[c]{@{}l@{}}0.523\end{tabular}
 & \begin{tabular}[c]{@{}l@{}}0.804\end{tabular} \\
 
 & standard CSM        
 & \begin{tabular}[c]{@{}l@{}}0.633\end{tabular}
 & \begin{tabular}[c]{@{}l@{}}0.505\end{tabular}
 & \begin{tabular}[c]{@{}l@{}}0.757\end{tabular} \\

 & weight01 CSM       
 & \begin{tabular}[c]{@{}l@{}}0.646\end{tabular}
 & \begin{tabular}[c]{@{}l@{}}0.514\end{tabular}
 & \begin{tabular}[c]{@{}l@{}}0.775\end{tabular} \\

 & \textbf{weightbeta CSM} 
 & \begin{tabular}[c]{@{}l@{}}\textbf{0.776}\end{tabular}
 & \begin{tabular}[c]{@{}l@{}}0.642\end{tabular}
 & \begin{tabular}[c]{@{}l@{}}0.876\end{tabular} \\
 
 & weightwald CSM       
 & \begin{tabular}[c]{@{}l@{}}0.615\end{tabular}
 & \begin{tabular}[c]{@{}l@{}}0.467\end{tabular}
 & \begin{tabular}[c]{@{}l@{}}0.745\end{tabular} \\ \hline

\multirow{6}{*}{ICI}
 & Euclidean distance
 & \begin{tabular}[c]{@{}l@{}}0.032\end{tabular}
 & \begin{tabular}[c]{@{}l@{}}0.017\end{tabular}
 & \begin{tabular}[c]{@{}l@{}}0.057\end{tabular} \\

 & Mahalanobis distance        
 & \begin{tabular}[c]{@{}l@{}}0.054\end{tabular}
 & \begin{tabular}[c]{@{}l@{}}0.024\end{tabular}
 & \begin{tabular}[c]{@{}l@{}}0.097\end{tabular} \\
 
 & \textbf{standard CSM}       
 & \begin{tabular}[c]{@{}l@{}}\textbf{0.029}\end{tabular}
 & \begin{tabular}[c]{@{}l@{}}0.012\end{tabular}
 & \begin{tabular}[c]{@{}l@{}}0.051\end{tabular} \\

 & weight01 CSM        
 & \begin{tabular}[c]{@{}l@{}}0.030\end{tabular}
 & \begin{tabular}[c]{@{}l@{}}0.013\end{tabular}
 & \begin{tabular}[c]{@{}l@{}}0.056\end{tabular} \\

 & weightbeta CSM      
 & \begin{tabular}[c]{@{}l@{}}0.057\end{tabular}
 & \begin{tabular}[c]{@{}l@{}}0.026\end{tabular}
 & \begin{tabular}[c]{@{}l@{}}0.092\end{tabular} \\
 
 & weightwald CSM      
 & \begin{tabular}[c]{@{}l@{}}0.031\end{tabular}
 & \begin{tabular}[c]{@{}l@{}}0.015\end{tabular}
 & \begin{tabular}[c]{@{}l@{}}0.058\end{tabular} \\ \hline

\multirow{6}{*}{BrS} 

 & Euclidean distance        
 & \begin{tabular}[c]{@{}l@{}}0.069\end{tabular}
 & \begin{tabular}[c]{@{}l@{}}0.054\end{tabular}
 & \begin{tabular}[c]{@{}l@{}}0.087\end{tabular} \\

 & Mahalanobis distance        
 & \begin{tabular}[c]{@{}l@{}}0.066\end{tabular}
 & \begin{tabular}[c]{@{}l@{}}0.047\end{tabular}
 & \begin{tabular}[c]{@{}l@{}}0.087\end{tabular} \\
 
 & standard CSM        
 & \begin{tabular}[c]{@{}l@{}}0.064\end{tabular}
 & \begin{tabular}[c]{@{}l@{}}0.049\end{tabular}
 & \begin{tabular}[c]{@{}l@{}}0.079\end{tabular} \\

 & weight01 CSM        
 & \begin{tabular}[c]{@{}l@{}}0.063\end{tabular}
 & \begin{tabular}[c]{@{}l@{}}0.049\end{tabular}
 & \begin{tabular}[c]{@{}l@{}}0.078\end{tabular} \\

 & \textbf{weightbeta CSM} 
 & \begin{tabular}[c]{@{}l@{}}\textbf{0.059}\end{tabular}
 & \begin{tabular}[c]{@{}l@{}}0.044\end{tabular}
 & \begin{tabular}[c]{@{}l@{}}0.077\end{tabular} \\
 
 & weightwald CSM 
 & \begin{tabular}[c]{@{}l@{}}0.066\end{tabular}
 & \begin{tabular}[c]{@{}l@{}}0.050\end{tabular}
 & \begin{tabular}[c]{@{}l@{}}0.083\end{tabular} \\ \hline

\end{tabular}
\label{tab:valcase2}
\end{table}

Figure~2 shows that Case 5, which has a reasonably high prevalence of 0.5 and a moderate signal, has a similar shape of AUROC against different subpopulation proportions compared to Case 2. Weight01 CSM and weightbeta CSM outperform standard CSM, weightwald CSM, and Euclidean distance, and in particular, the discriminative ability of weightbeta CSM has a dramatic improvement compared to all other similarity measures. In Case 5, best discrimination is achieved by training the model on the top 0.1, or 10\%, of the most similar participants in the available data. Calibration has an interesting outcome compared to Case 2. First, the best calibration is achieved by using weightwald CSM when the subpopulation proportion is 10\%. Furthermore, when the dataset has a reasonably high prevalence, we see a unique shape in the plot of ICI against different subpopulation proportions for weightbeta CSM. It has an inverse parabolic shape, where the calibration at a subpopulation proportion of 0.1 is much better than that of standard CSM and weight01 CSM. With BrS, which considers both discrimination and calibration, we also observe a different pattern compared to Case 2. For both weighted and unweighted CSM methods, lower subpopulation proportion results in better BrS, and the minimum value is achieved with weightbeta CSM at a subpopulation proportion of 0.1. Overall, weightbeta CSM and weight01 CSM outperform all other measures of similarity when we train the data on a more targeted and homogeneous subset of the available training data.

Tuning for the optimal subpopulation proportion that will result in the best BrS demonstrates that with Case 5, using the top 10\% most similar patients in the training data leads to the best predictive performance of the predictive model. Using this value, we can validate the performance of the model using the holdout validation set, where the results can be found in Table~2. From the tuning process using the TrTe data, we note that while using weightbeta CSM does not lead to the best calibration in PPM, gains in discrimination are much bigger when the subpopulation proportion is 0.1; then using 0.1 as our chosen subpopulation proportion from the tuning process, we note that even with calibration being considered, weightbeta CSM outperforms standard CSM across all three performance metrics. 

\begin{table}[h] 
\caption{Results from the validation study for a reasonably high prevalence, moderate signal (Case 5). The columns ``Lower'' and ``Upper'' represent the 2.5\% BCa and 97.5\% BCa bootstrap interval estimates, respectively, and the winning $M_p$ used to calculate the performance metrics is $0.1$ from the tuning process.}
\centering
\begin{tabular}{|l|l|l|l|l|}
\hline
\multicolumn{1}{|c|}{Metric} & \multicolumn{1}{c|}{Similarity metric} & \multicolumn{1}{c|}{Est} & \multicolumn{1}{c|}{Lower} & \multicolumn{1}{c|}{Upper} \\ \hline

\multirow{6}{*}{AUROC}
 & Euclidean distance        
 & \begin{tabular}[c]{@{}l@{}}0.801\end{tabular}
 & \begin{tabular}[c]{@{}l@{}}0.752\end{tabular}
 & \begin{tabular}[c]{@{}l@{}}0.844\end{tabular} \\

 & Mahalanobis distance   
 & \begin{tabular}[c]{@{}l@{}}0.909\end{tabular}
 & \begin{tabular}[c]{@{}l@{}}0.874\end{tabular}
 & \begin{tabular}[c]{@{}l@{}}0.935\end{tabular} \\
 
 & standard CSM        
 & \begin{tabular}[c]{@{}l@{}}0.879\end{tabular}
 & \begin{tabular}[c]{@{}l@{}}0.843\end{tabular}
 & \begin{tabular}[c]{@{}l@{}}0.914\end{tabular} \\

 & weight01 CSM   
 & \begin{tabular}[c]{@{}l@{}}0.932\end{tabular}
 & \begin{tabular}[c]{@{}l@{}}0.888\end{tabular}
 & \begin{tabular}[c]{@{}l@{}}0.956\end{tabular} \\

 & \textbf{weightbeta CSM} 
 & \begin{tabular}[c]{@{}l@{}}\textbf{0.942}\end{tabular}
 & \begin{tabular}[c]{@{}l@{}}0.913\end{tabular}
 & \begin{tabular}[c]{@{}l@{}}0.966\end{tabular} \\ 
 
 & weightwald CSM   
 & \begin{tabular}[c]{@{}l@{}}0.873\end{tabular}
 & \begin{tabular}[c]{@{}l@{}}0.826\end{tabular}
 & \begin{tabular}[c]{@{}l@{}}0.918\end{tabular} \\\hline

\multirow{6}{*}{ICI}

 & Euclidean distance        
 & \begin{tabular}[c]{@{}l@{}}0.124\end{tabular}
 & \begin{tabular}[c]{@{}l@{}}0.084\end{tabular}
 & \begin{tabular}[c]{@{}l@{}}0.171\end{tabular} \\

 & Mahalanobis distance   
 & \begin{tabular}[c]{@{}l@{}}0.147\end{tabular}
 & \begin{tabular}[c]{@{}l@{}}0.082\end{tabular}
 & \begin{tabular}[c]{@{}l@{}}0.202\end{tabular} \\
 
 & standard CSM        
 & \begin{tabular}[c]{@{}l@{}}0.100\end{tabular}
 & \begin{tabular}[c]{@{}l@{}}0.055\end{tabular}
 & \begin{tabular}[c]{@{}l@{}}0.151\end{tabular} \\

 & weight01 CSM   
 & \begin{tabular}[c]{@{}l@{}}0.104\end{tabular}
 & \begin{tabular}[c]{@{}l@{}}0.065\end{tabular}
 & \begin{tabular}[c]{@{}l@{}}0.150\end{tabular} \\

 & weightbeta CSM
 & \begin{tabular}[c]{@{}l@{}}0.084\end{tabular}
 & \begin{tabular}[c]{@{}l@{}}0.058\end{tabular}
 & \begin{tabular}[c]{@{}l@{}}0.121\end{tabular} \\ 
 
 & \textbf{weightwald CSM}   
 & \begin{tabular}[c]{@{}l@{}}\textbf{0.065}\end{tabular}
 & \begin{tabular}[c]{@{}l@{}}0.034\end{tabular}
 & \begin{tabular}[c]{@{}l@{}}0.105\end{tabular} \\\hline

\multirow{6}{*}{BrS}

 & Euclidean distance        
 & \begin{tabular}[c]{@{}l@{}}0.200\end{tabular}
 & \begin{tabular}[c]{@{}l@{}}0.171\end{tabular}
 & \begin{tabular}[c]{@{}l@{}}0.233\end{tabular} \\

 & Mahalanobis distance   
 & \begin{tabular}[c]{@{}l@{}}0.152\end{tabular}
 & \begin{tabular}[c]{@{}l@{}}0.121\end{tabular}
 & \begin{tabular}[c]{@{}l@{}}0.189\end{tabular} \\ 
 
 & standard CSM        
 & \begin{tabular}[c]{@{}l@{}}0.150\end{tabular}
 & \begin{tabular}[c]{@{}l@{}}0.121\end{tabular}
 & \begin{tabular}[c]{@{}l@{}}0.180\end{tabular} \\

 & weight01 CSM   
 & \begin{tabular}[c]{@{}l@{}}0.116\end{tabular}
 & \begin{tabular}[c]{@{}l@{}}0.087\end{tabular}
 & \begin{tabular}[c]{@{}l@{}}0.149\end{tabular} \\

 & \textbf{weightbeta CSM} 
 & \begin{tabular}[c]{@{}l@{}}\textbf{0.087}\end{tabular}
 & \begin{tabular}[c]{@{}l@{}}0.063\end{tabular}
 & \begin{tabular}[c]{@{}l@{}}0.116\end{tabular} \\ 
 
 & weightwald CSM   
 & \begin{tabular}[c]{@{}l@{}}0.147\end{tabular}
 & \begin{tabular}[c]{@{}l@{}}0.116\end{tabular}
 & \begin{tabular}[c]{@{}l@{}}0.179\end{tabular} \\ \hline

\end{tabular}
\label{tab:valcase5}
\end{table}

Table~3 displays the relative difference in calculating similarity using weightbeta CSM versus standard CSM for each of the four metrics across different scenario cases, obtained by applying the tuned $M_p$ values to the holdout validation set. Relative difference is calculated using the following equation:
\begin{equation} \label{RelDiff}
    RelDiff_{wbeta} = \frac{|metric_{wbetaCSM} - metric_{standardCSM}|}{metric_{standardCSM}}.
\end{equation}
The tuned $M_p$ values applied to the validation set to obtain the performance metrics are shown in the final column. We can observe that generally, as the signal grows stronger - i.e. the beta values used to generate the outcome get larger - the relative difference in the measures increases. This is as expected, since the outcome is generated using only ten predictors and our proposed weightbeta CSM aims to identify and use these predictors to measure the similarity between the participants instead of using all predictors available. Thus, if the signal is stronger, then we expect the calculated similarity to be more dissimilar between assigning weights on these selected features versus using all 20 predictors equally to evaluate the similarity between participants. 

\begin{table}[] 
\caption{Relative difference in the performance measures between using weightbeta CSM and standard CSM, and the tuned $M_p$ value for each simulation case.}
\resizebox{\textwidth}{!}{
\begin{tabular}{|l|l|l|l|l|l|}
\hline
\multicolumn{1}{|c|}{Case} & \multicolumn{1}{c|}{$\textbf{AUROC}_{\Delta rel}$} & \multicolumn{1}{c|}{$\textbf{AUPRC}_{\Delta rel}$} & \multicolumn{1}{c|}{$\textbf{ICI}_{\Delta rel}$} & \multicolumn{1}{c|}{$\textbf{BrS}_{\Delta rel}$} & \multicolumn{1}{c|}{Tuned $M_p$} \\ \hline

Case 1: Prev 0.1, weak  &  & \begin{tabular}[c]{@{}l@{}}0.2050\end{tabular} & \begin{tabular}[c]{@{}l@{}}0.9989\end{tabular} & \begin{tabular}[c]{@{}l@{}}0.0249\end{tabular} & 0.6 \\ \hline

Case 2: Prev 0.1, mod   &  & \begin{tabular}[c]{@{}l@{}}0.2254\end{tabular} & \begin{tabular}[c]{@{}l@{}}1.0137\end{tabular} & \begin{tabular}[c]{@{}l@{}}0.0806\end{tabular} & 0.6 \\ \hline

Case 3: Prev 0.1, strong &  & \begin{tabular}[c]{@{}l@{}}0.2429\end{tabular} & \begin{tabular}[c]{@{}l@{}}1.0221\end{tabular} & \begin{tabular}[c]{@{}l@{}}0.1027\end{tabular} & 0.6 \\ \hline

Case 4: Prev 0.5, weak  & \begin{tabular}[c]{@{}l@{}}0.0738\end{tabular} &  & \begin{tabular}[c]{@{}l@{}}0.0141\end{tabular} & \begin{tabular}[c]{@{}l@{}}0.2244\end{tabular} & 0.1 \\ \hline

Case 5: Prev 0.5, mod   & \begin{tabular}[c]{@{}l@{}}0.0712\end{tabular} &  & \begin{tabular}[c]{@{}l@{}}0.1409\end{tabular} & \begin{tabular}[c]{@{}l@{}}0.4223\end{tabular} & 0.1 \\ \hline

Case 6: Prev 0.5, strong & \begin{tabular}[c]{@{}l@{}}0.0660\end{tabular} &  & \begin{tabular}[c]{@{}l@{}}0.1815\end{tabular} & \begin{tabular}[c]{@{}l@{}}0.4332\end{tabular} & 0.1 \\ \hline

\end{tabular}
}
\label{tab:metricimprovement}
\end{table}

Table~4 displays the relative difference in calculating similarity using weight01 CSM versus standard CSM for each of the four metrics across different scenario cases, obtained by applying the tuned $M_p$ values to the holdout validation set. Relative difference is calculated using the following equation:
\begin{equation} \label{RelDiff}
    RelDiff_{w01} = \frac{|metric_{w01CSM} - metric_{standardCSM}|}{metric_{standardCSM}}.
\end{equation}

\begin{table}[] 
\caption{Relative difference in the performance measures between using weight01 CSM and standard CSM, and the tuned $M_p$ value for each simulation case.}
\resizebox{\textwidth}{!}{
\begin{tabular}{|l|l|l|l|l|l|}
\hline
\multicolumn{1}{|c|}{Case} & \multicolumn{1}{c|}{$\textbf{AUROC}_{\Delta rel}$} & \multicolumn{1}{c|}{$\textbf{AUPRC}_{\Delta rel}$} & \multicolumn{1}{c|}{$\textbf{ICI}_{\Delta rel}$} & \multicolumn{1}{c|}{$\textbf{BrS}_{\Delta rel}$} & \multicolumn{1}{c|}{Tuned $M_p$} \\ \hline

Case 1: Prev 0.1, weak  &  & \begin{tabular}[c]{@{}l@{}}0.0218\end{tabular} & \begin{tabular}[c]{@{}l@{}}0.0946\end{tabular} & \begin{tabular}[c]{@{}l@{}}0.0070\end{tabular} & 0.6 \\ \hline

Case 2: Prev 0.1, mod   &  & \begin{tabular}[c]{@{}l@{}}0.0211\end{tabular} & \begin{tabular}[c]{@{}l@{}}0.0542\end{tabular} & \begin{tabular}[c]{@{}l@{}}0.0155\end{tabular} & 0.6 \\ \hline

Case 3: Prev 0.1, strong &  & \begin{tabular}[c]{@{}l@{}}0.0416\end{tabular} & \begin{tabular}[c]{@{}l@{}}0.1733\end{tabular} & \begin{tabular}[c]{@{}l@{}}0.0322\end{tabular} & 0.6 \\ \hline

Case 4: Prev 0.5, weak  & \begin{tabular}[c]{@{}l@{}}0.0393\end{tabular} &  & \begin{tabular}[c]{@{}l@{}}0.0891\end{tabular} & \begin{tabular}[c]{@{}l@{}}0.1100\end{tabular} & 0.1 \\ \hline

Case 5: Prev 0.5, mod   & \begin{tabular}[c]{@{}l@{}}0.0601\end{tabular} &  & \begin{tabular}[c]{@{}l@{}}0.0631\end{tabular} & \begin{tabular}[c]{@{}l@{}}0.2288\end{tabular} & 0.1 \\ \hline

Case 6: Prev 0.5, strong & \begin{tabular}[c]{@{}l@{}}0.0514\end{tabular} &  & \begin{tabular}[c]{@{}l@{}}0.0120\end{tabular} & \begin{tabular}[c]{@{}l@{}}0.2233\end{tabular} & 0.1 \\ \hline

\end{tabular}
}
\label{tab:metricimprovement01}
\end{table}

There are several subtle differences between Table~3 and Table~4. First, the relative differences in the performance metrics are generally larger in Table~3. Second, unlike Table~3, which reports the relative differences between weightbeta CSM and standard CSM, Table~4 does not show a clear monotonic increase in relative differences for weight01 CSM versus standard CSM as the signal strength increases. Although stronger signals tend to produce larger relative differences than weaker signals, the pattern is not strictly monotonic. We speculate that this lack of monotonicity may be due to the way $M_p$ is tuned. $M_p$ is determined as the value that will give the smallest BrS across all three CSM methods. Since weightbeta CSM outperforms weight01 CSM and standard CSM measured by BrS from our simulation study, $M_p$ is selected based on the predictive performance of the model using weightbeta CSM.

\section{Data Analysis} \label{sec:data analysis}

We apply our proposed algorithms in Section~2.1 and~2.2 to a publicly available intensive care unit database, called the eICU Collaborative Research Database. This multi-center critical care database was made available by Philips Healthcare with the MIT Laboratory for Computational Physiology and is publicly available on the PhysioNet repository \citep{pollard2018eicu}. The eICU database holds a large amount of data associated with over 200,000 patient intensive care unit (ICU) stays, containing information about vital sign measurements, nurses' notes, diagnoses, treatment decisions, patient demographic information and more from a combination of critical care units throughout the continental United States in 2014 and 2015. To get an independent dataset with each patient having only one record, we focus on only the first ICU stay of the first hospital stay, where identifiable, of the patients who were in the ICU for at least 24 hours, have no missing APACHE (i.e. APACHE IVa) score - a predictor for mortality that incorporates patient information collected within the first 24 hours of the ICU stay, including both demographic information about the patient and physiologic data \citep{zimmerman2006acute} - and have no missing vital sign data in the first 24 hours of their ICU stay. This workflow is outlined in Figure~3.

\begin{figure}[ht]
    \centering
    \includegraphics[width=0.6\linewidth]{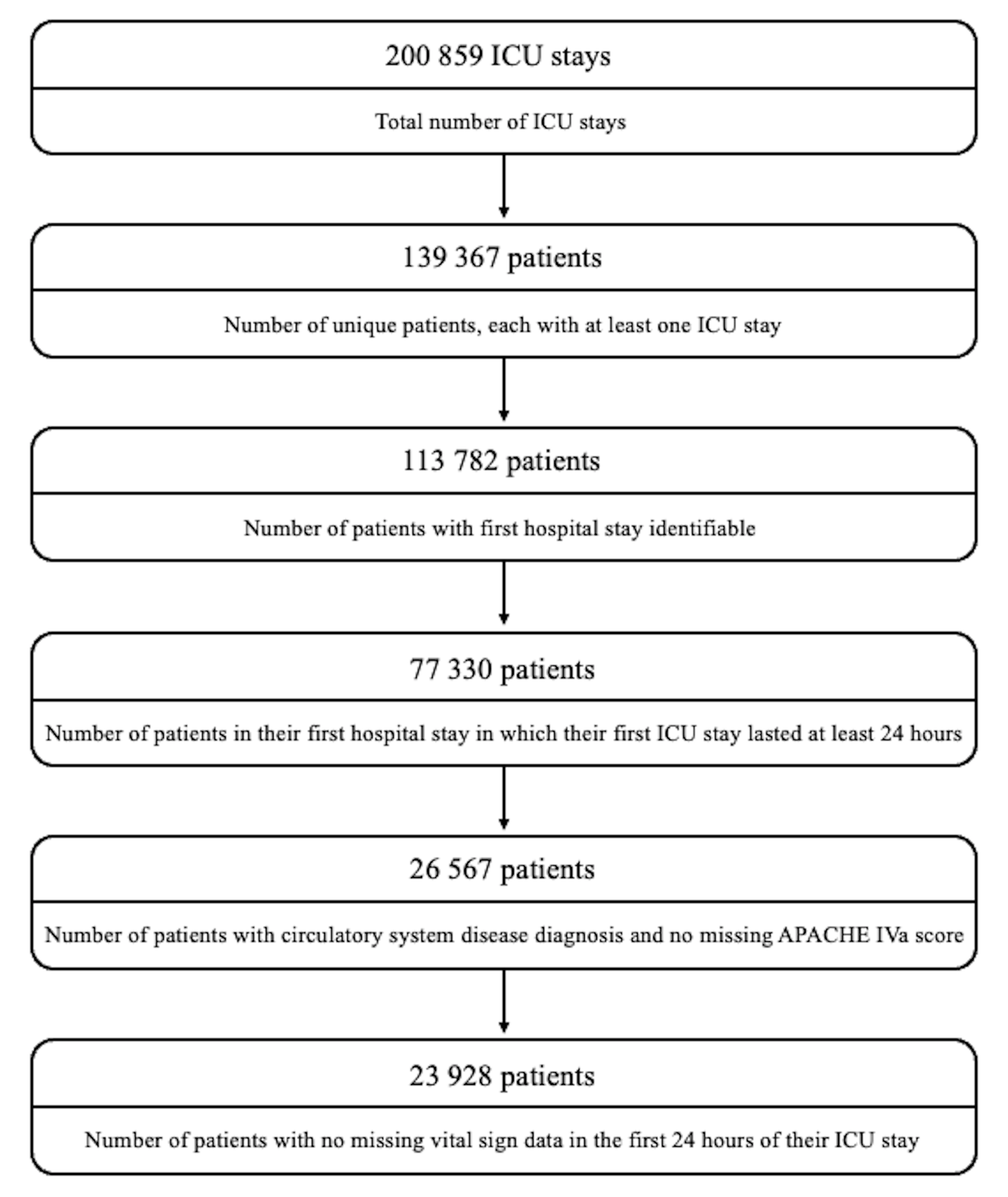}
    \caption{Flow chart of cohort selection from the eICU database for data analysis.}
    \label{fig:eicuflow}
\end{figure}

For the predictor variables, we consider age, gender, densely collected longitudinal vital signs - including oxygen saturation (Sao2), respiration rate, and heart rate - for the first 24 hours after ICU admission, and APACHE IVa score. We use the hospital discharge status of dead or alive as the binary outcome variable. To represent most of the variation in the longitudinal vital signs measurements for the first 24 hours of the ICU stay, we use functional principal component analysis (FPCA), a multivariate analysis technique that extracts information from functional data, such as longitudinal data \citep{wang2016functional}. It reduces the dimension of data with a large number of interrelated variables and transforms the data to a new set of variables, called functional principal components, while retaining as much of the total variation as possible \citep{ullah2013applications}. The principal components are uncorrelated and can be ordered so that the first few will retain most of the variation present in the original variable. This is implemented using the Principal Analysis by Conditional Estimation (PACE) algorithm in the R package \textbf{fdapace} \citep{fdapace} where the expansion for the longitudinal measurement $Y_i(t)$ with only the first $r$ eigenfunctions ($\hat{\phi}_k(t), k = 1, 2, ..., r)$ is 
\begin{equation}
\hat{Y}_i^{r}(t) = \hat{\mu}(t) + \sum^{r}_{k=1} \hat{\xi}_{ik} \hat{\phi}_k(t) ,   
\end{equation}
where $\hat{\mu}(t)$ is the estimated mean function and $\hat{\xi}_{ik}$ are the FPCs. In this work, we choose three FPCs for each biomarker as in each case, this explained at least 95\% of the variation in the variable.

\subsection{Results} \label{sec:data analysis results}

We apply PPM to the eICU data, with $0.1, 0.2, 0.3, 0.4, 0.5, 0.6, 0.7, 0.8, 0.9,$ and $1$ considered for the $M_p$ values, and calculate the patient similarity using standard CSM, weight01 CSM, and weightbeta CSM. We consider these three similarity metrics only in our data analysis because we have shown in the simulation study that weight01 CSM and weightbeta CSM outperform other common measures of similarity. 80\% of the available dataset is used for training and testing the model, using a 10-fold CV repeated ten times across different subpopulation proportions (Figure~4). We measure discrimination using AUPRC, since this dataset has a low prevalence of 0.11. The general pattern is similar to the simulation study, where as the subpopulation proportion increases, AUPRC decreases for weight01 CSM and weightbeta CSM. For a standard CSM that uses all the available predictors, AUPRC peaks when the subpopulation proportion is 0.6 and decreases thereafter. Best AUPRC is achieved with weightbeta CSM at a subpopulation proportion of 0.2. For calibration measured with ICI, using weight01 CSM almost always outperforms using standard CSM or weightbeta CSM. Best calibration is achieved with weight01 CSM when the subpopulation proportion is 0.2. For BrS, we also see the same result that calculating patient similarity using weight01 CSM and weightbeta CSM outperform standard CSM, with the lowest, and thus best BrS value, achieved with weightbeta CSM when the subpopulation proportion is 0.2. 

From AUPRC, ICI, and BrS, the best predictive performance of the model can be achieved for a subpopulation proportion of 0.2. Using the holdout validation set, we evaluate the performance of the model when we train the predictive model on the top 20\% most similar participants for the participant of interest and calculate the performance metrics. The results are found in Table~5. Compared to the simulation study, the difference in the predictive performance of the model when using standard CSM and weightbeta CSM is much smaller. However, the estimates and the bias-corrected and accelerated bootstrap confidence intervals show that generally, AUPRC is higher and ICI and BrS are lower (i.e., all in the preferred direction) when we use weightbeta CSM instead of weight01 CSM or standard CSM to implement PPM and predict the outcome. 

\begin{figure}[htbp]
\centering

\begin{minipage}[t]{0.49\textwidth}
    \centering
    \includegraphics[width=\textwidth]{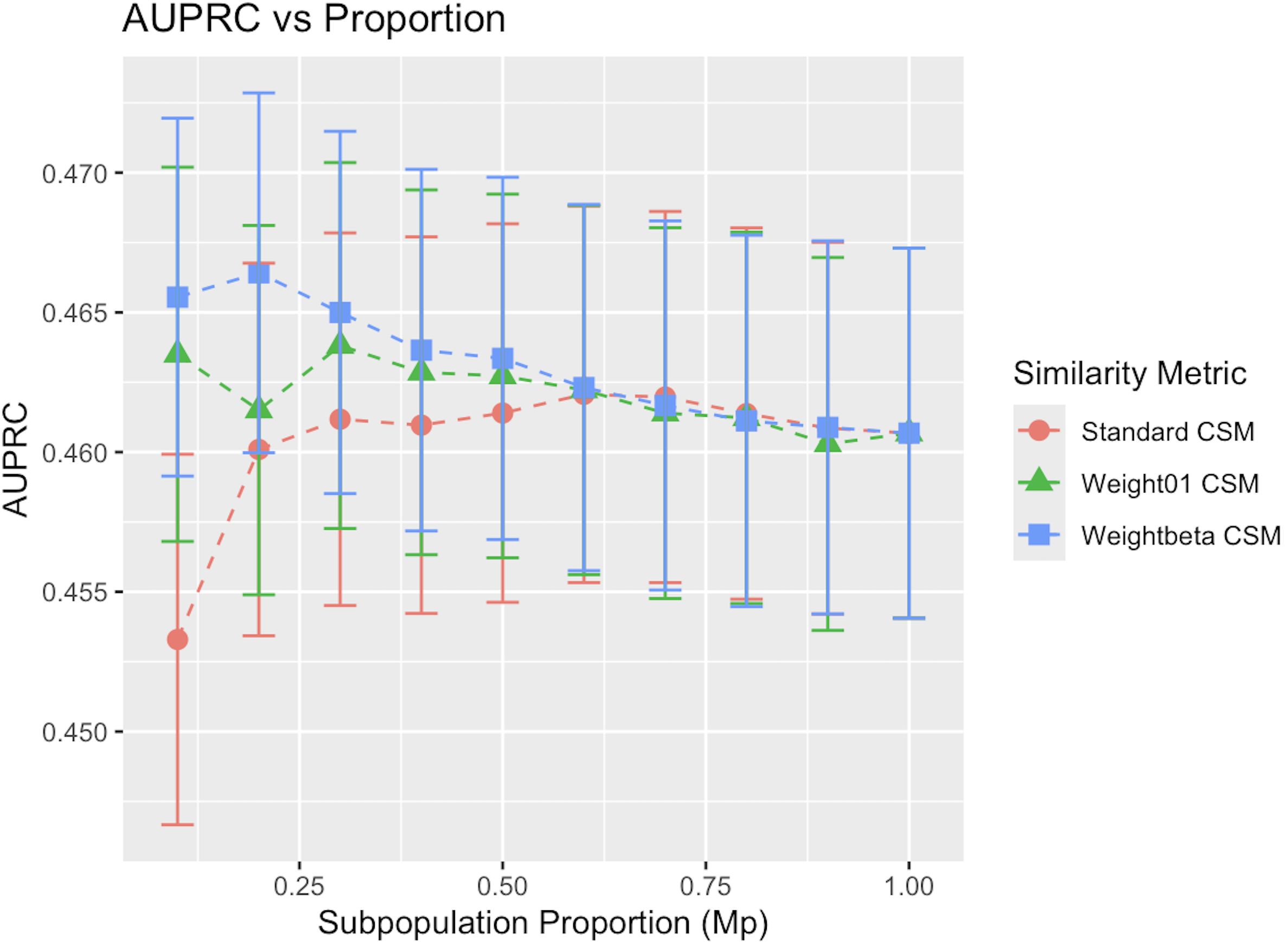}
    \label{fig:plot1}
\end{minipage}
\hfill
\begin{minipage}[t]{0.49\textwidth}
    \centering
    \includegraphics[width=\textwidth]{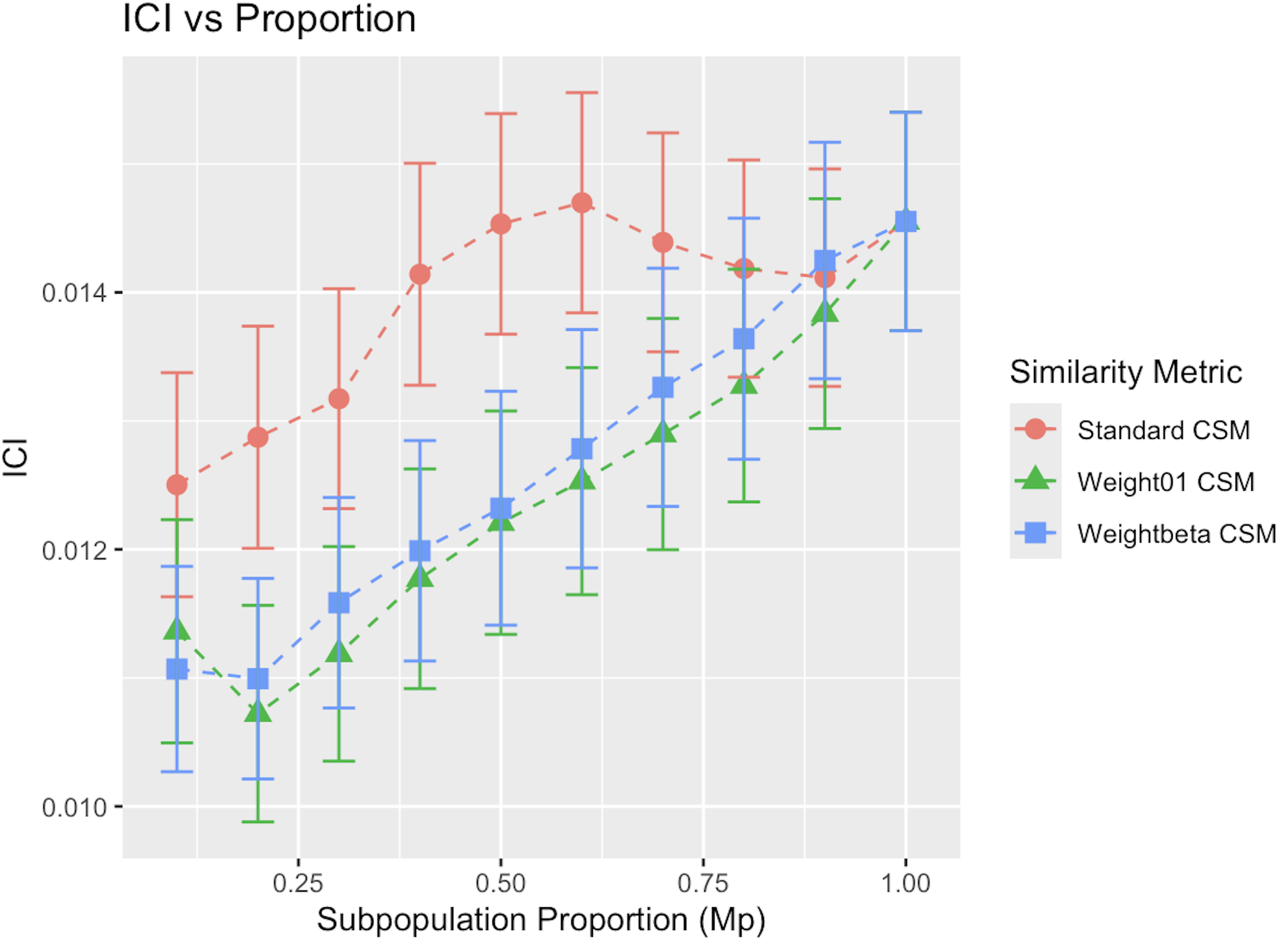}
    \label{fig:plot2}
\end{minipage}

\vspace{0.5cm}

\begin{minipage}[t]{0.49\textwidth}
    \centering
    \includegraphics[width=\linewidth]{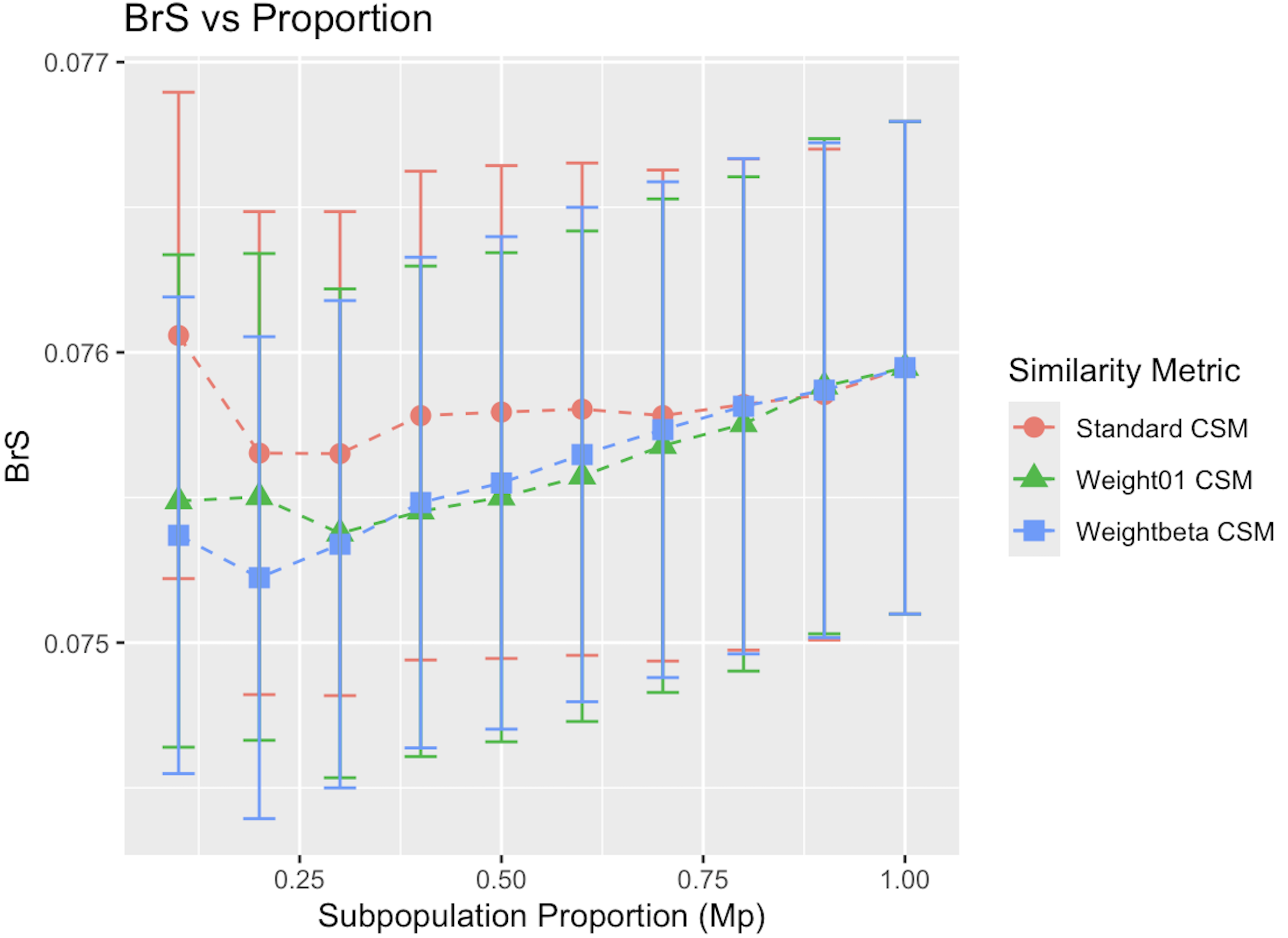}
\end{minipage}\hfill%
\begin{minipage}[t]{0.5\textwidth}
\centering
\vspace{0pt}
\small   
\end{minipage}

\caption{Performance metrics across different subpopulation proportions for the three similarity measures using the ICU dataset.}
\label{fig:eicutuning}
\end{figure}

\begin{table}[] 
\caption{Results from the validation study for data analysis using the ICU dataset. The columns ``Lower'' and ``Upper'' represent the 2.5\% BCa and 97.5\% BCa bootstrap interval estimates, respectively, and the winning $M_p$ used to calculate the performance metrics is $0.2$ from the tuning process.}
\centering
\begin{tabular}{|l|l|l|l|l|}
\hline
\multicolumn{1}{|c|}{Metric} & \multicolumn{1}{c|}{Similarity Metric} & \multicolumn{1}{c|}{Est} & \multicolumn{1}{c|}{Lower} & \multicolumn{1}{c|}{Upper} \\ \hline

\multirow{3}{*}{AUPRC}
 & standard CSM        
 & \begin{tabular}[c]{@{}l@{}}0.4562\end{tabular}
 & \begin{tabular}[c]{@{}l@{}}0.3603\end{tabular}
 & \begin{tabular}[c]{@{}l@{}}0.5596\end{tabular} \\

 & weight01 CSM   
 & \begin{tabular}[c]{@{}l@{}}0.4612\end{tabular}
 & \begin{tabular}[c]{@{}l@{}}0.3606\end{tabular}
 & \begin{tabular}[c]{@{}l@{}}0.5704\end{tabular} \\

 & \textbf{weightbeta CSM} 
 & \begin{tabular}[c]{@{}l@{}}\textbf{0.4648}\end{tabular}
 & \begin{tabular}[c]{@{}l@{}}0.3633\end{tabular}
 & \begin{tabular}[c]{@{}l@{}}0.5633\end{tabular} \\ \hline

\multirow{3}{*}{ICI}
 & standard CSM        
 & \begin{tabular}[c]{@{}l@{}}0.0195\end{tabular}
 & \begin{tabular}[c]{@{}l@{}}0.0095\end{tabular}
 & \begin{tabular}[c]{@{}l@{}}0.0340\end{tabular} \\

 & \textbf{weight01 CSM}   
 & \begin{tabular}[c]{@{}l@{}}\textbf{0.0192}\end{tabular}
 & \begin{tabular}[c]{@{}l@{}}0.0094\end{tabular}
 & \begin{tabular}[c]{@{}l@{}}0.0350\end{tabular} \\

 & weightbeta CSM      
 & \begin{tabular}[c]{@{}l@{}}0.0196\end{tabular}
 & \begin{tabular}[c]{@{}l@{}}0.0095\end{tabular}
 & \begin{tabular}[c]{@{}l@{}}0.0350\end{tabular} \\ \hline

\multirow{3}{*}{BrS}
 & standard CSM        
 & \begin{tabular}[c]{@{}l@{}}0.0724\end{tabular}
 & \begin{tabular}[c]{@{}l@{}}0.0606\end{tabular}
 & \begin{tabular}[c]{@{}l@{}}0.0852\end{tabular} \\

 & weight01 CSM   
 & \begin{tabular}[c]{@{}l@{}}0.0719\end{tabular}
 & \begin{tabular}[c]{@{}l@{}}0.0607\end{tabular}
 & \begin{tabular}[c]{@{}l@{}}0.0849\end{tabular} \\

 & \textbf{weightbeta CSM} 
 & \begin{tabular}[c]{@{}l@{}}\textbf{0.0716}\end{tabular}
 & \begin{tabular}[c]{@{}l@{}}0.0600\end{tabular}
 & \begin{tabular}[c]{@{}l@{}}0.0844\end{tabular} \\ \hline

\end{tabular}
\label{tab:valeicu}
\end{table}

\section{Discussion} \label{sec:discussion}

We have shown that we can improve the predictive performance of the model using PPM when we implement our new proposed methods to calculate the patient similarity. In PPM, we aim to identify a subset of the data that is the most similar to the participant of interest so that we can train a model for each person in the testing data on a more tailored and homogeneous dataset and improve the prediction. Therefore, measuring similarity between participants to identify this subset of the dataset is crucial. CSM is a popular method to do so, but it can be biased and lead to suboptimal results \citep{li2013distance}; thus, researchers have introduced a weighted version of CSM where features are assigned different levels of importance in similarity calculation. However, in current literature, the weights are derived using the properties of the data, such as feature frequency, or from expert knowledge that reflects the importance of features in a given context (for example, CSM with weights derived using Wu \& Palmer measure \citep{wu1994verb} or the Lin measure \citep{lin1998information} require previous knowledge about the hierarchical structure of the features and their relationships.) We introduce a new weighted similarity metric that can be used to determine the weights without any prior knowledge about the types of predictors and remains effective even when some predictors undergo unknown transformations before influencing the outcome. We introduce two similarity metrics using relaxed adaptive group lasso: weight01 CSM, which is a weighted cosine similarity metric that assigns a weight of 0 or 1 to each available predictor, and weightbeta CSM, which assigns weights to each predictor based on the beta estimates after implementing relaxed adaptive group lasso. 

Using the simulation study with generated data and data analysis using the ICU dataset, we demonstrate that we can greatly improve discrimination, and calibration in some cases, of a prediction model when implementing our proposed method for similarity calculation. The simulation study considered six cases: two levels of prevalence, and three levels of the strength of association between the predictors and the outcome. We consider AUROC, AUPRC, ICI and BrS to evaluate the model performance. When the prevalence is low at 0.1, training the model on a subset of the available training data improves discrimination when using weightbeta CSM or weight01 CSM compared to standard CSM and other existing similarity metrics. However, when considering calibration, standard CSM outperforms weight01 CSM or weightbeta CSM. When we consider both discrimination and calibration using BrS, weightbeta CSM and weight01 CSM outperform other similarity metrics at all considered subpopulation proportions, with BrS minimized using weightbeta CSM and a subpopulation proportion of 0.6. When the prevalence of the data is high at 0.5, discrimination shows the same pattern as the dataset with low prevalence, where using weightbeta CSM and weight01 CSM results in better prediction measures with AUROC maximized at subpopulation proportion of 0.1. However, there is a difference in calibration results, where it is better to use weightbeta CSM compared to standard CSM when the subpopulation proportion is 0.1. Therefore, BrS can be minimized when the subpopulation proportion is 0.1 using weightbeta CSM. When considering the three levels of association between the predictors and the outcome, we see that generally, as the strength increases, the relative difference in the performance metric when implementing PPM with weightbeta CSM versus standard CSM increases as well. In other words, when the signal is stronger, we expect PPM using weightbeta CSM to have better discrimination.

Data analysis using ICU shows similar results. Considering AUPRC, ICI, and BrS to evaluate the discrimination and calibration of the prediction model, using weightbeta CSM to implement PPM results in better predictive performance compared to using standard CSM. Highest AUPRC, lowest ICI, and lowest BrS values are achieved when training the prediction model on the top 20\% of the most similar data using weightbeta CSM. The biggest difference in the results of the simulation study and data analysis is in calibration. In the simulation study, calibration is usually worse when using weight01 CSM or weightbeta CSM instead of standard CSM for different subpopulation proportions when the prevalence is 0.1. Calibration is better when using weightbeta CSM only at the subpopulation proportion of 0.1 when the prevalence is 0.5. Furthermore, ICI is the smallest across all similarity metrics when the subpopulation proportion is 0.7 or 0.8. However, with the ICU data, ICI is the smallest when the subpopulation proportion is 0.2, and using weightbeta CSM consistently performs better than using standard CSM across all subpopulation proportions. We suspect that this may be due to different characteristics of the datasets, and further investigation is required to define how certain characteristics of the data, such as the number of predictors, the association between these predictors and the outcome, associations among predictors, and the types of model fit to the data, can affect these results.

There are several areas of work for future study. First, more investigation with simulations should be conducted to be able to explain the different patterns in the calibration results. Calibration is very sensitive to how the dataset is generated, as we have seen from the different conclusions in the simulation results and the data analysis. We also plan to evaluate different calibration adjustment methods to improve the overall predictive performance of the model, such as intercept and slope recalibration \citep{van2019calibration}. Furthermore, we want to explore implementing a nested CV algorithm to create a nested CV interval estimate introduced by Bates et al. around the performance measures to create intervals that do not underestimate the variance \citep{bates2023cross}. Another limitation is not confirming our results using a sophisticated machine learning model, in large part due to the computational burden of the methodology. Implementing PPM is extremely heavy computationally, because we have to fit a new model for each participant in the test data and repeat this process for every fold when using repeated CV to tune for the optimal subpopulation proportion. In our simulation study and data analysis, we use a generalized linear model, so the process is manageable; using a machine learning model, such as random forest, for example, would certainly increase the computation time. However, when we implement PPM with a simple generalized linear model using weightbeta CSM to calculate the similarity between participants, we see an improvement in the model's predictive performance at a subpopulation proportion less than 1 compared to training a random forest model on the entire training data. The next step would be to implement PPM with a random forest model to investigate how the performance metrics change with varying subpopulation proportions and similarity metrics. Finally, we assume linearity in the predictors through the link function. Although this keeps the model simple and interpretable, it may not capture more complex relationships among the variables. Future work could explore more flexible modelling approaches that allow nonlinear effects and interactions between predictors. Variable-importance measures could also be used to better understand the relative contribution of individual predictors and guide model development.

\newpage

\bibliographystyle{plainnat}  
\bibliography{References}     

\newpage

\appendix

\section{Simulation Results}
We provide the simulation results for Case 1, Case 3, Case 4, and Case 6 from the simulation study. While these results are mentioned in Section 3.2, the detailed output from the tuning process on TrTe data and holdout Validation data for each case is shown below. When the prevalence is low at 0.1, we show AUPRC, ICI, and BrS plotted against various $M_p$ values, with the performance measures validated using holdout validation data and the winning $M_p$ of 0.6. When the prevalence is reasonably high at 0.5, we show AUROC, ICI, and BrS plotted against various $M_p$ values, with the performance measures validated using holdout validation data and the winning $M_p$ of 0.1.

\begin{figure}[htbp]
\centering

\begin{minipage}[t]{0.49\textwidth}
    \centering
    \includegraphics[width=\textwidth]{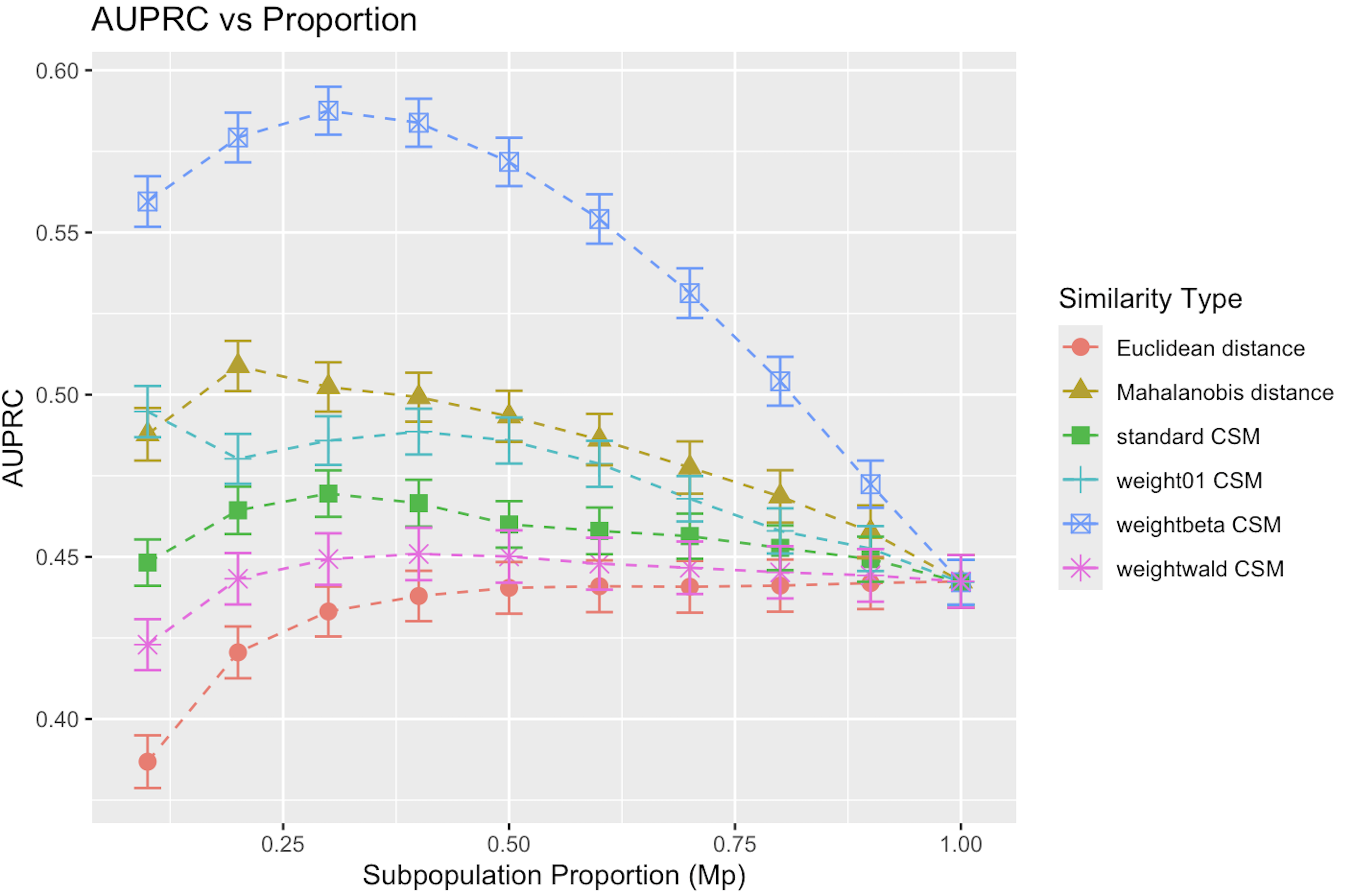}
    \label{fig:plot1}
\end{minipage}
\hfill
\begin{minipage}[t]{0.49\textwidth}
    \centering
    \includegraphics[width=\textwidth]{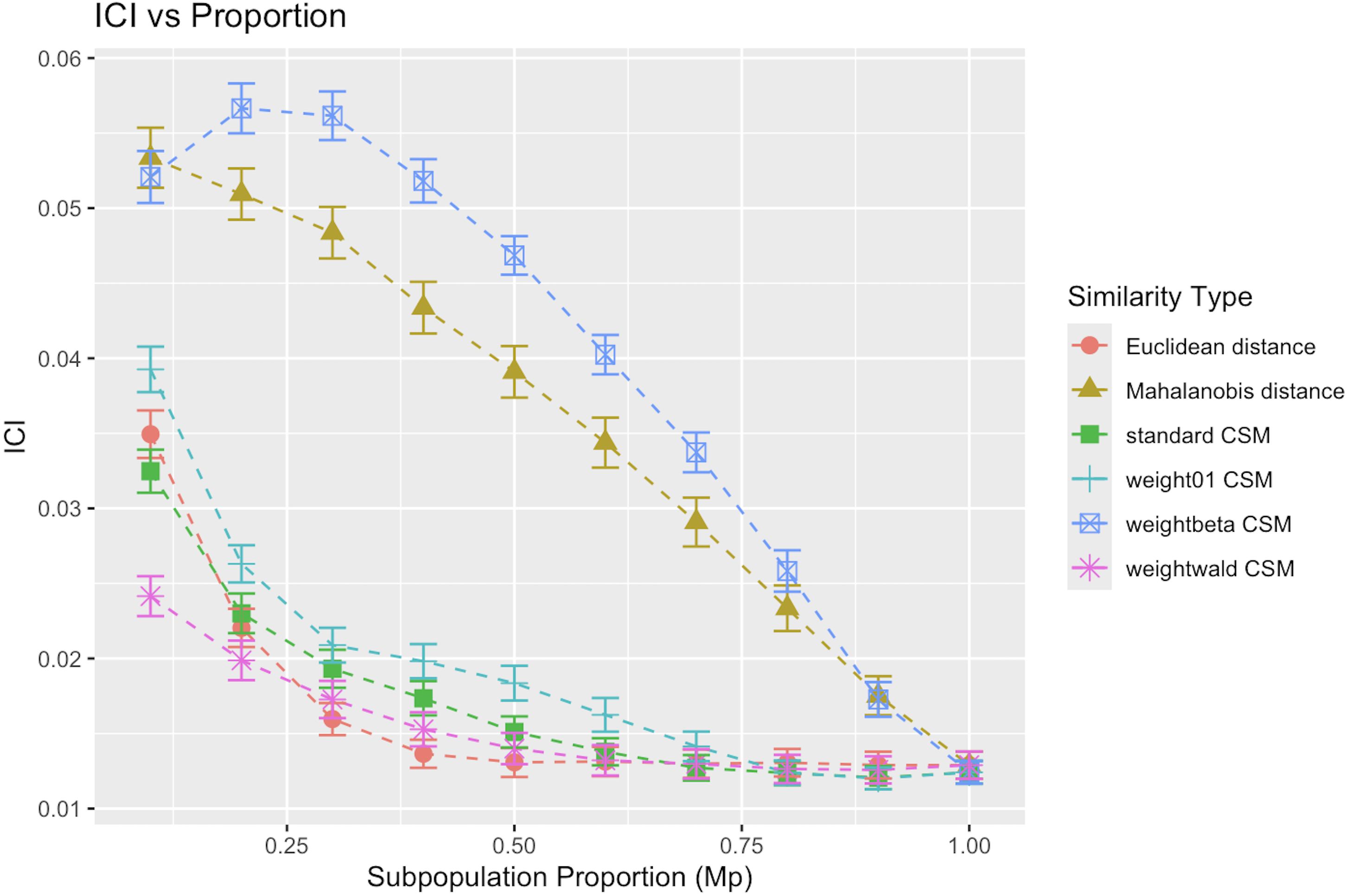}
    \label{fig:plot2}
\end{minipage}

\vspace{0.5cm}

\begin{minipage}[t]{0.49\textwidth}
    \centering
    \includegraphics[width=\textwidth]{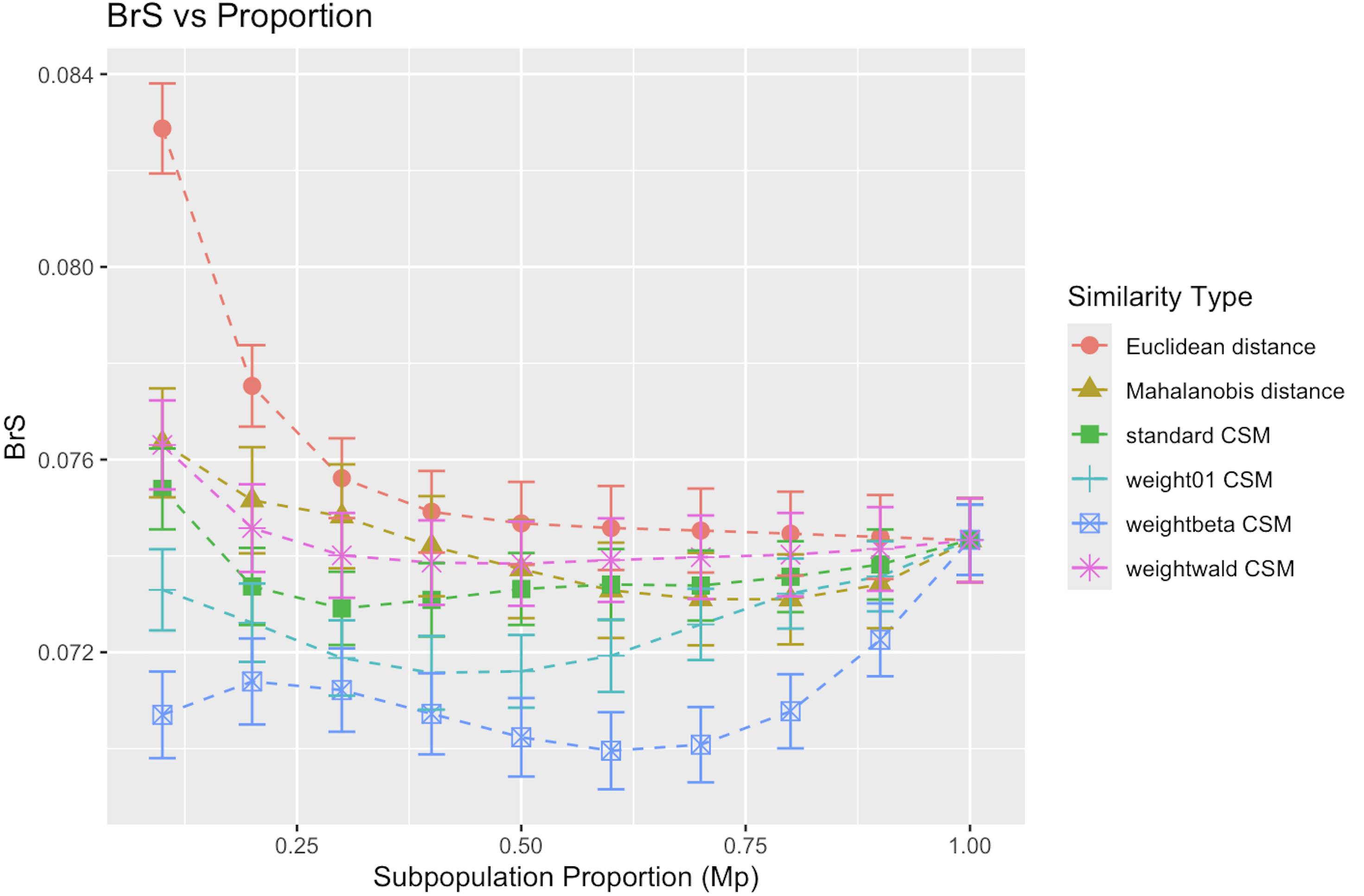}
    \label{fig:plot1}
\end{minipage}
\hfill
\begin{minipage}[t]{0.49\textwidth}
    \centering
    \includegraphics[width=\textwidth]{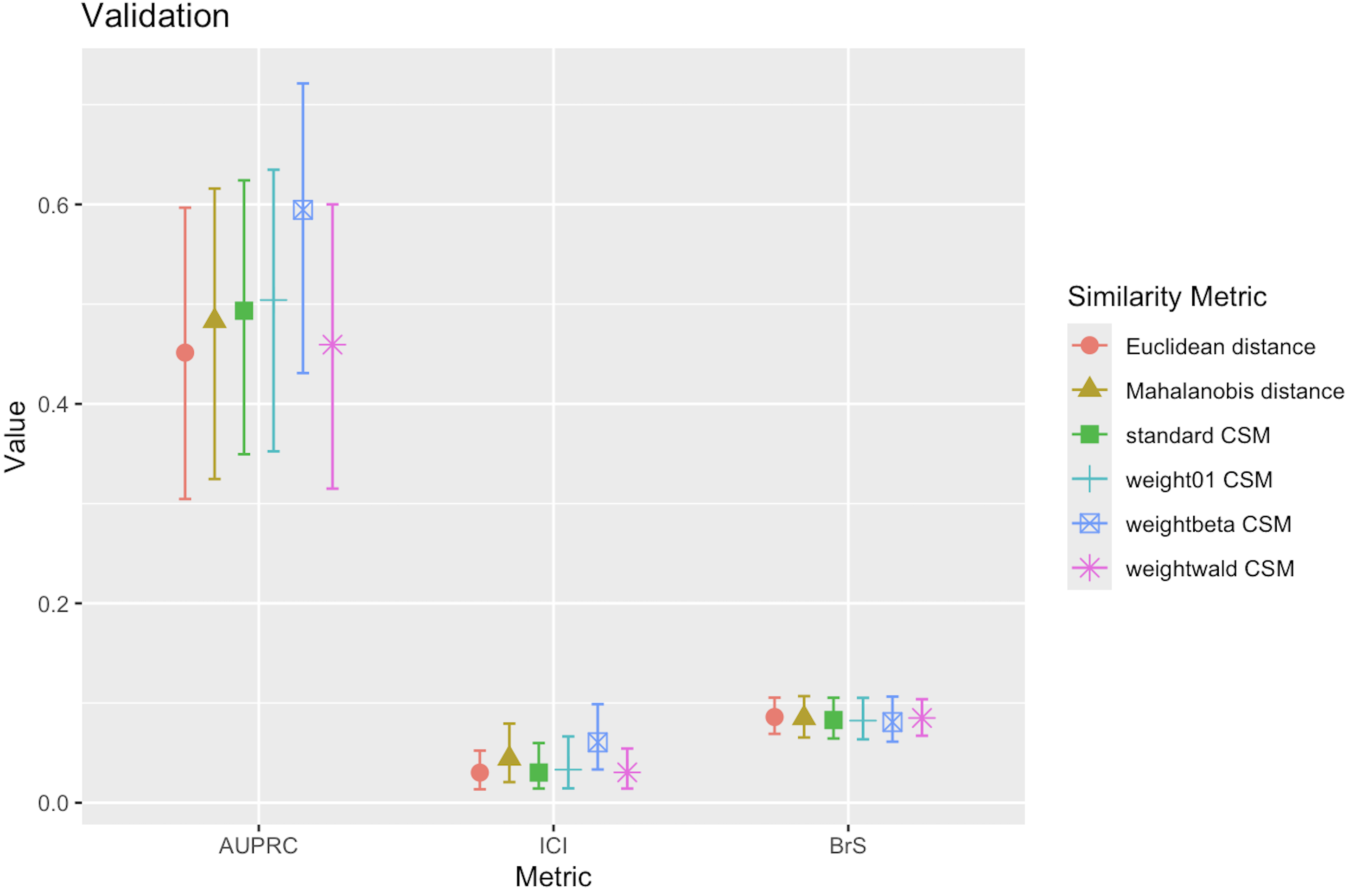}
    \label{fig:plot2}
\end{minipage}

\caption{Performance metrics across different subpopulation proportions for the six similarity measures. The dataset has low prevalence with a weak signal (Case 1).}
\label{fig:4plotscase1}
\end{figure}

\begin{figure}[htbp]
\centering

\begin{minipage}[t]{0.49\textwidth}
    \centering
    \includegraphics[width=\textwidth]{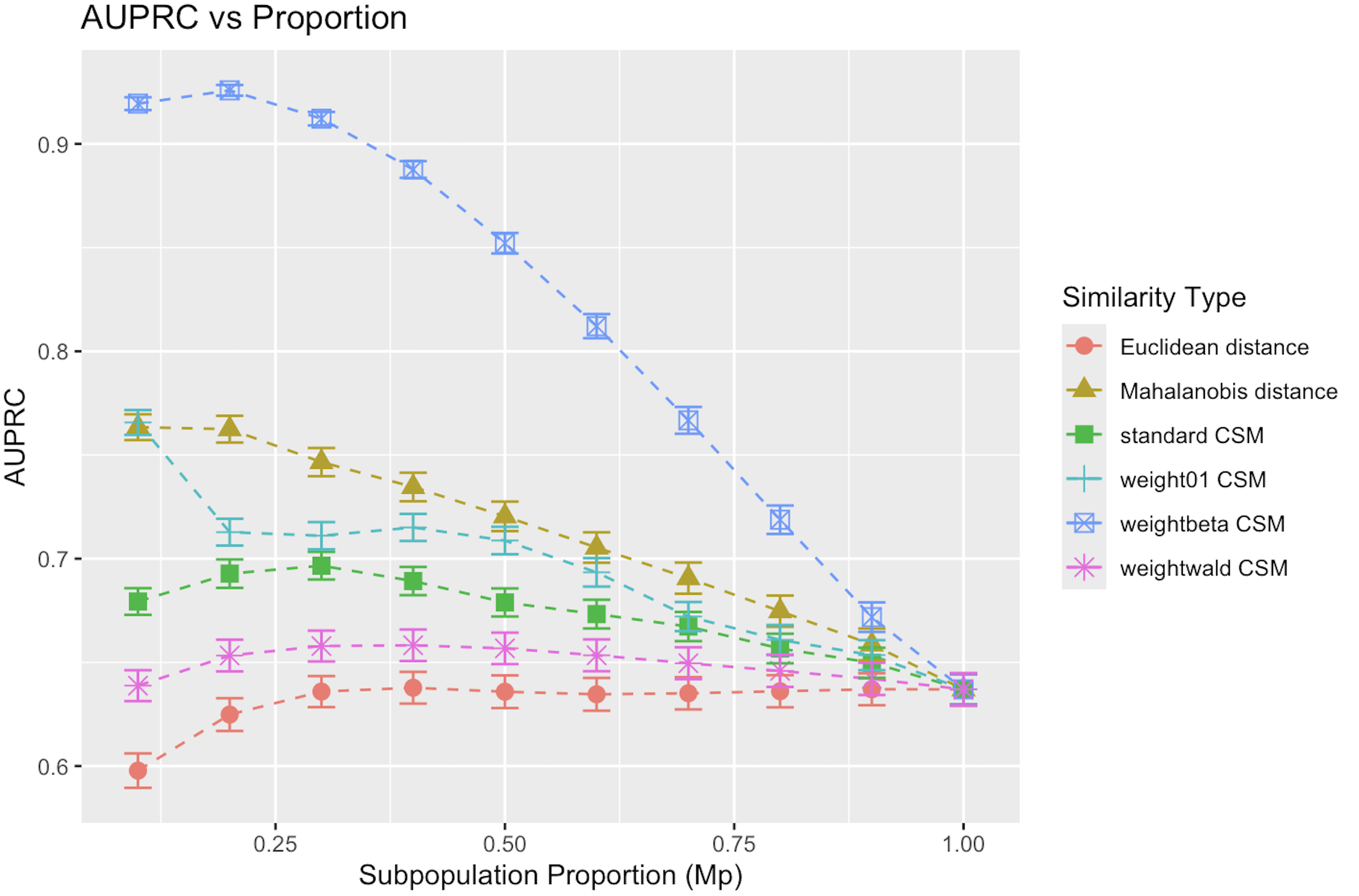}
    \label{fig:plot1}
\end{minipage}
\hfill
\begin{minipage}[t]{0.49\textwidth}
    \centering
    \includegraphics[width=\textwidth]{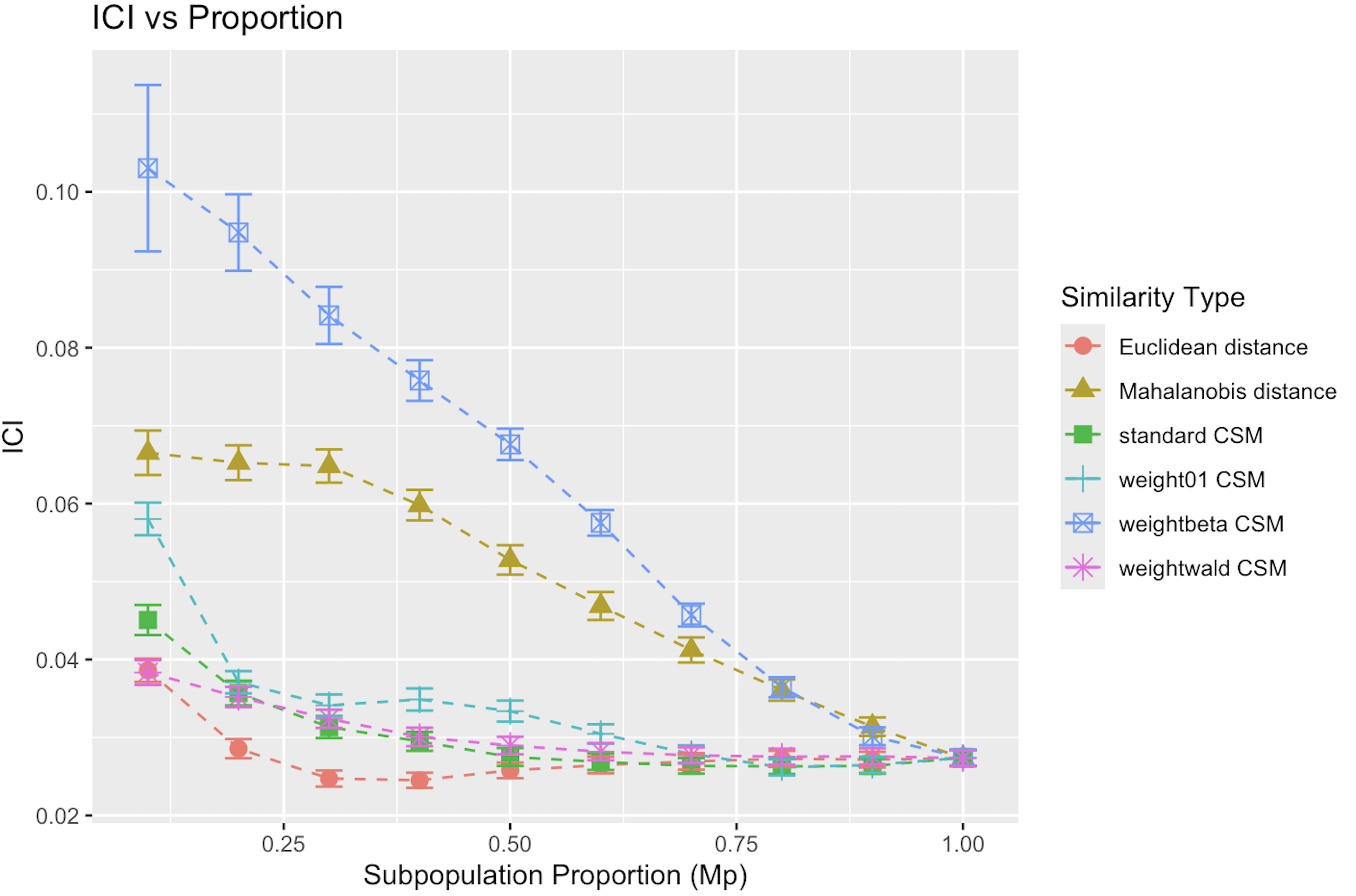}
    \label{fig:plot2}
\end{minipage}

\vspace{0.5cm}

\begin{minipage}[t]{0.49\textwidth}
    \centering
    \includegraphics[width=\textwidth]{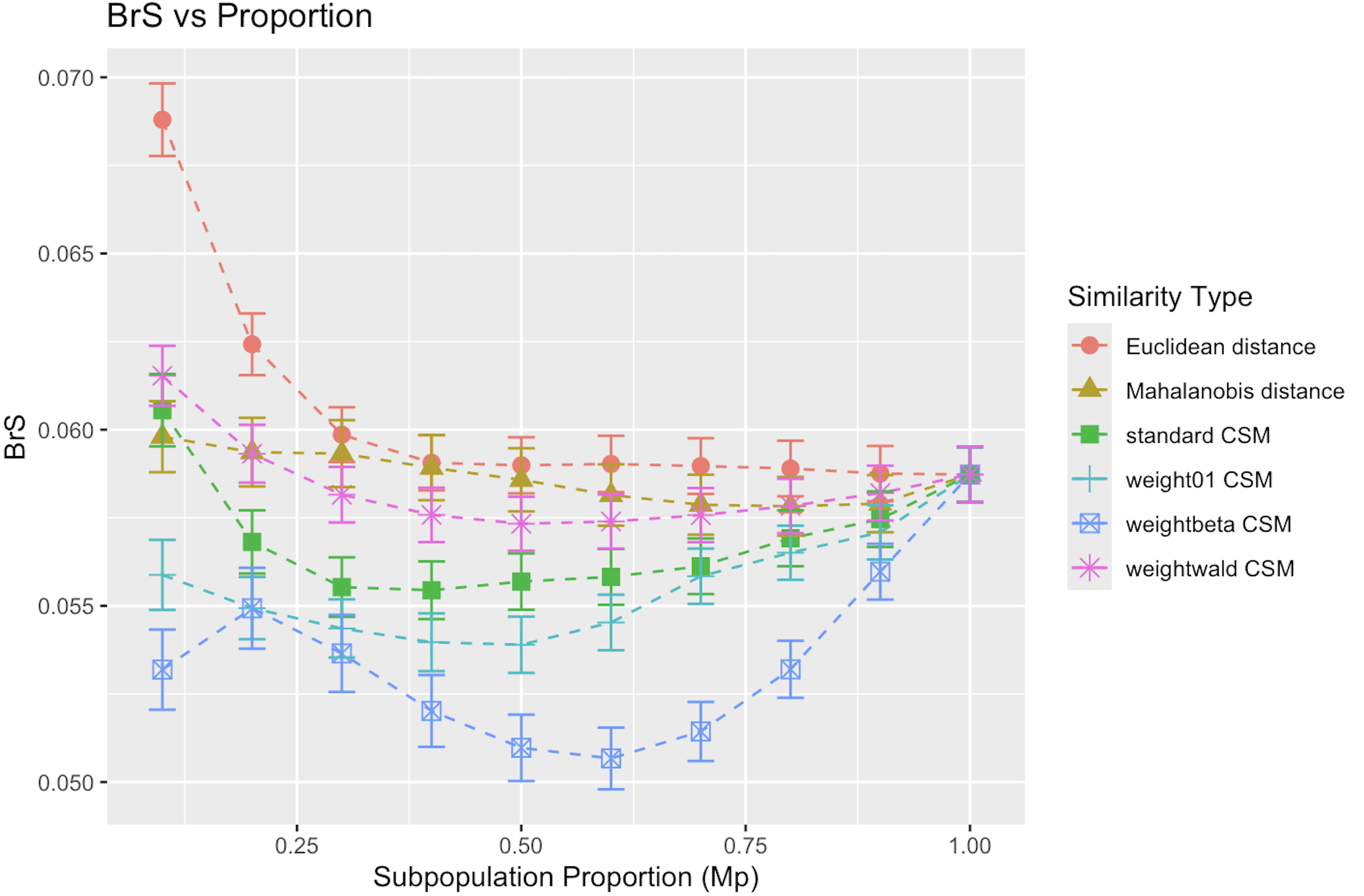}
    \label{fig:plot1}
\end{minipage}
\hfill
\begin{minipage}[t]{0.49\textwidth}
    \centering
    \includegraphics[width=\textwidth]{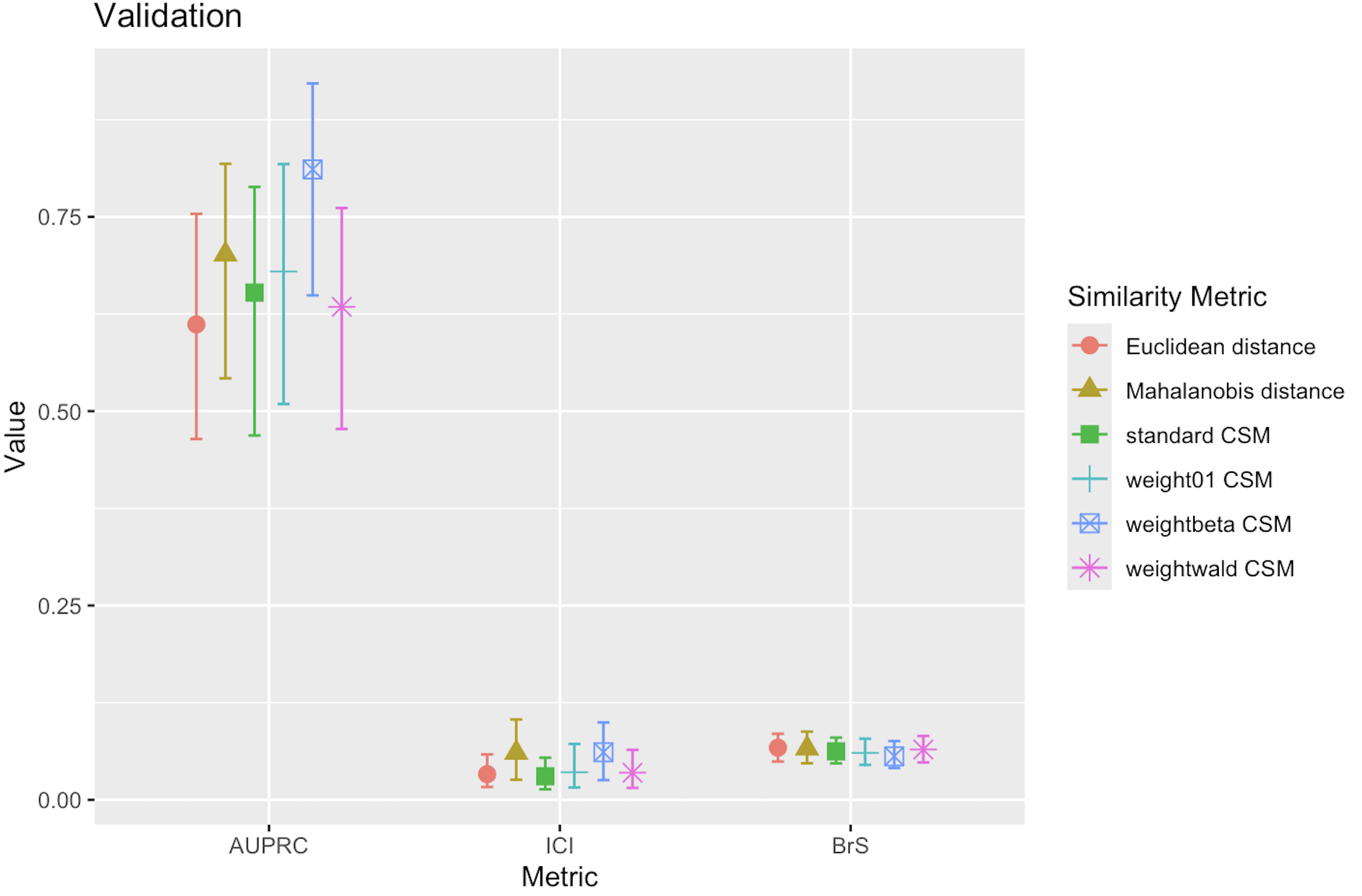}
    \label{fig:plot2}
\end{minipage}

\caption{Performance metrics across different subpopulation proportions for the six similarity measures. The dataset has low prevalence with a strong signal (Case 3).}
\label{fig:4plotscase3}
\end{figure}

\begin{figure}[htbp]
\centering

\begin{minipage}[t]{0.49\textwidth}
    \centering
    \includegraphics[width=\textwidth]{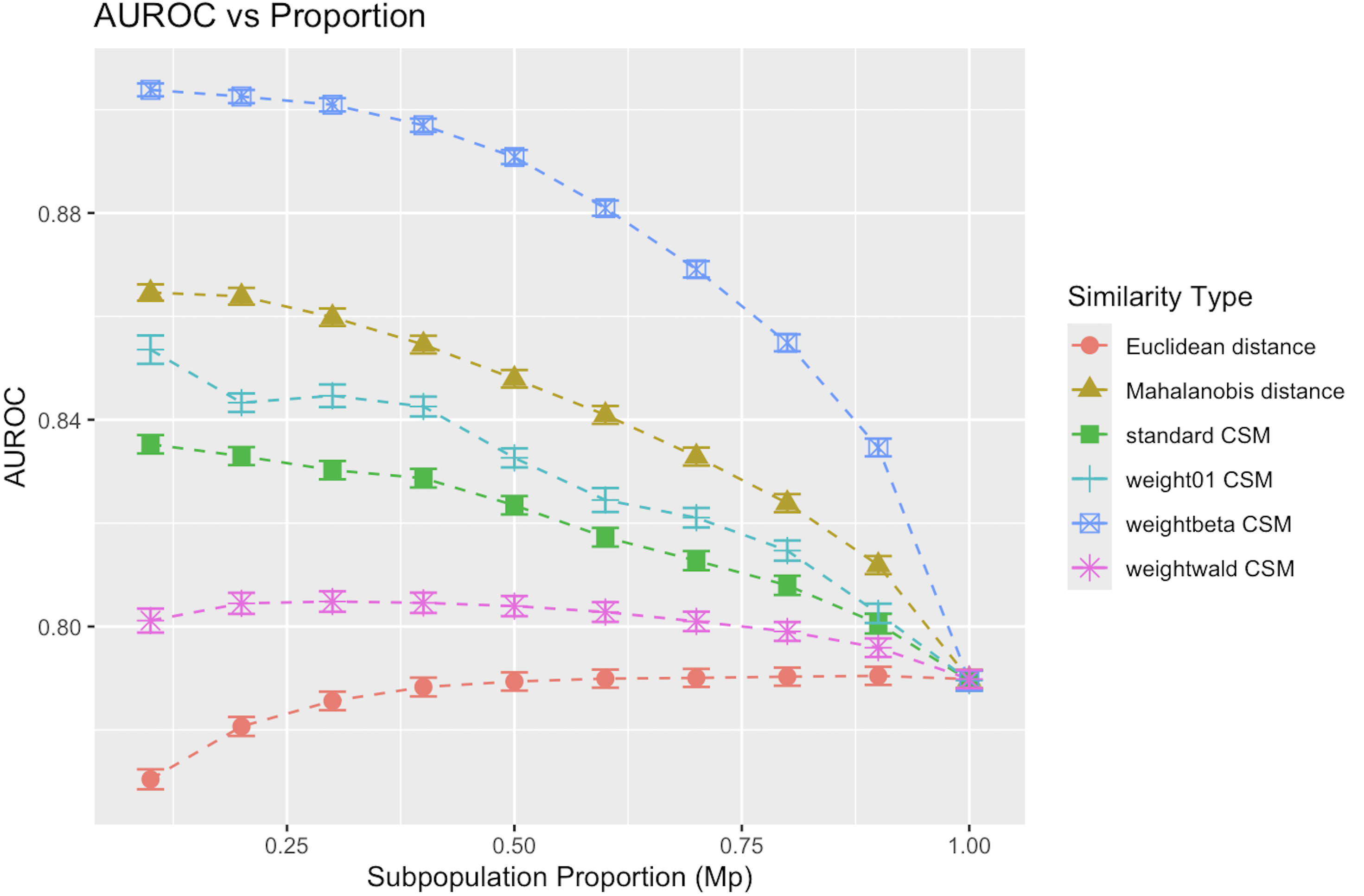}
    \label{fig:plot1}
\end{minipage}
\hfill
\begin{minipage}[t]{0.49\textwidth}
    \centering
    \includegraphics[width=\textwidth]{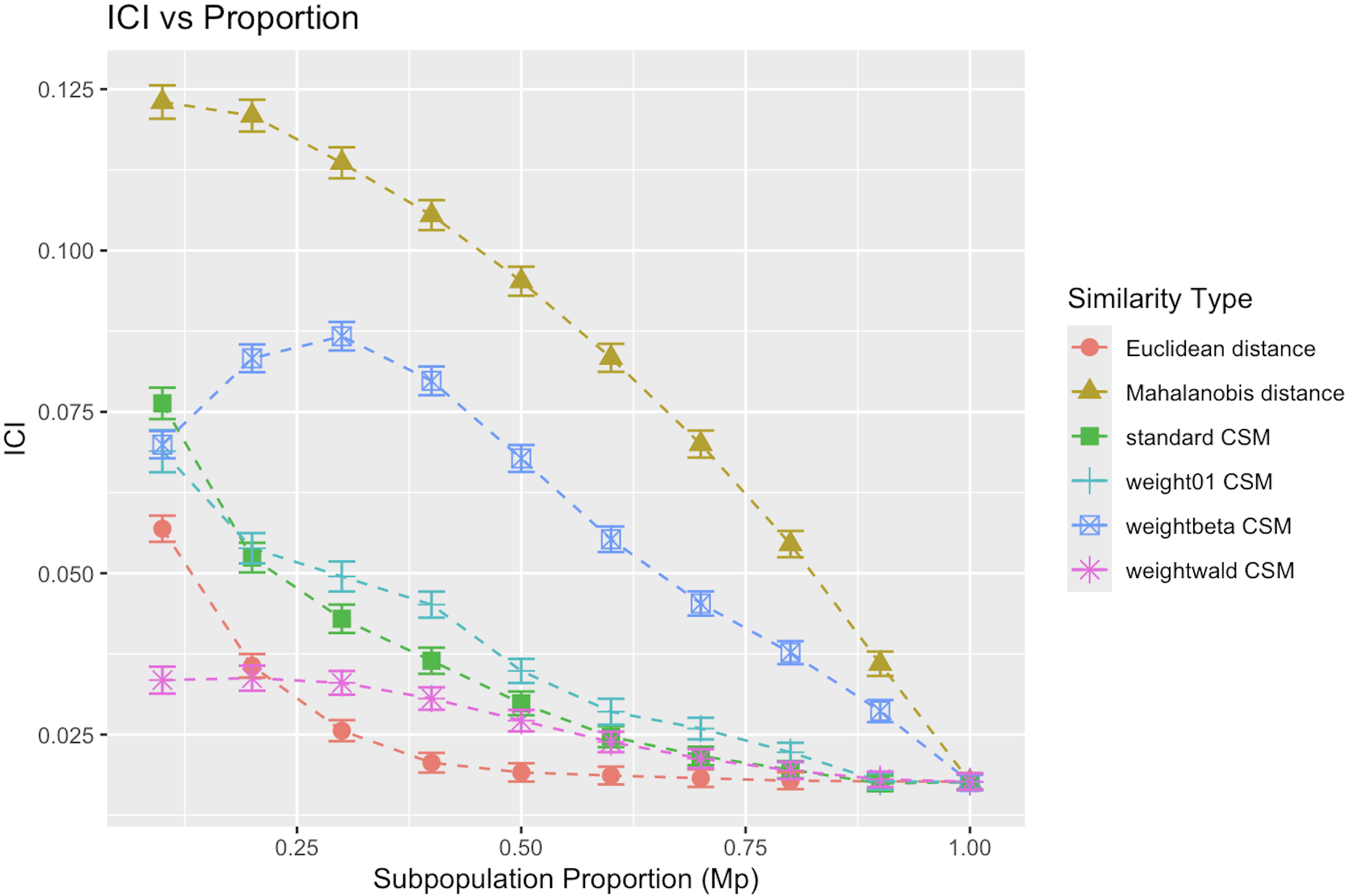}
    \label{fig:plot2}
\end{minipage}

\vspace{0.5cm}

\begin{minipage}[t]{0.49\textwidth}
    \centering
    \includegraphics[width=\textwidth]{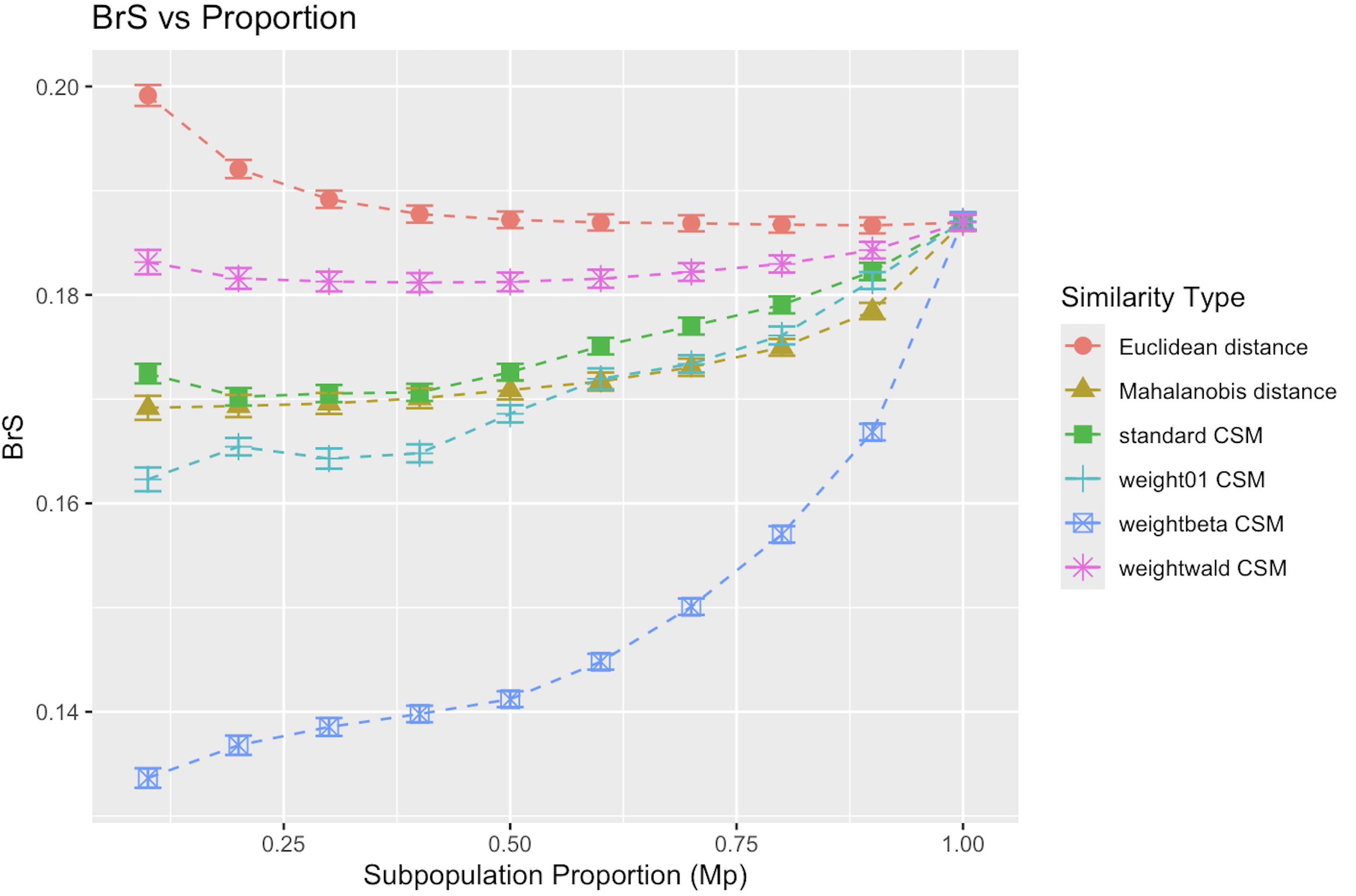}
    \label{fig:plot1}
\end{minipage}
\hfill
\begin{minipage}[t]{0.49\textwidth}
    \centering
    \includegraphics[width=\textwidth]{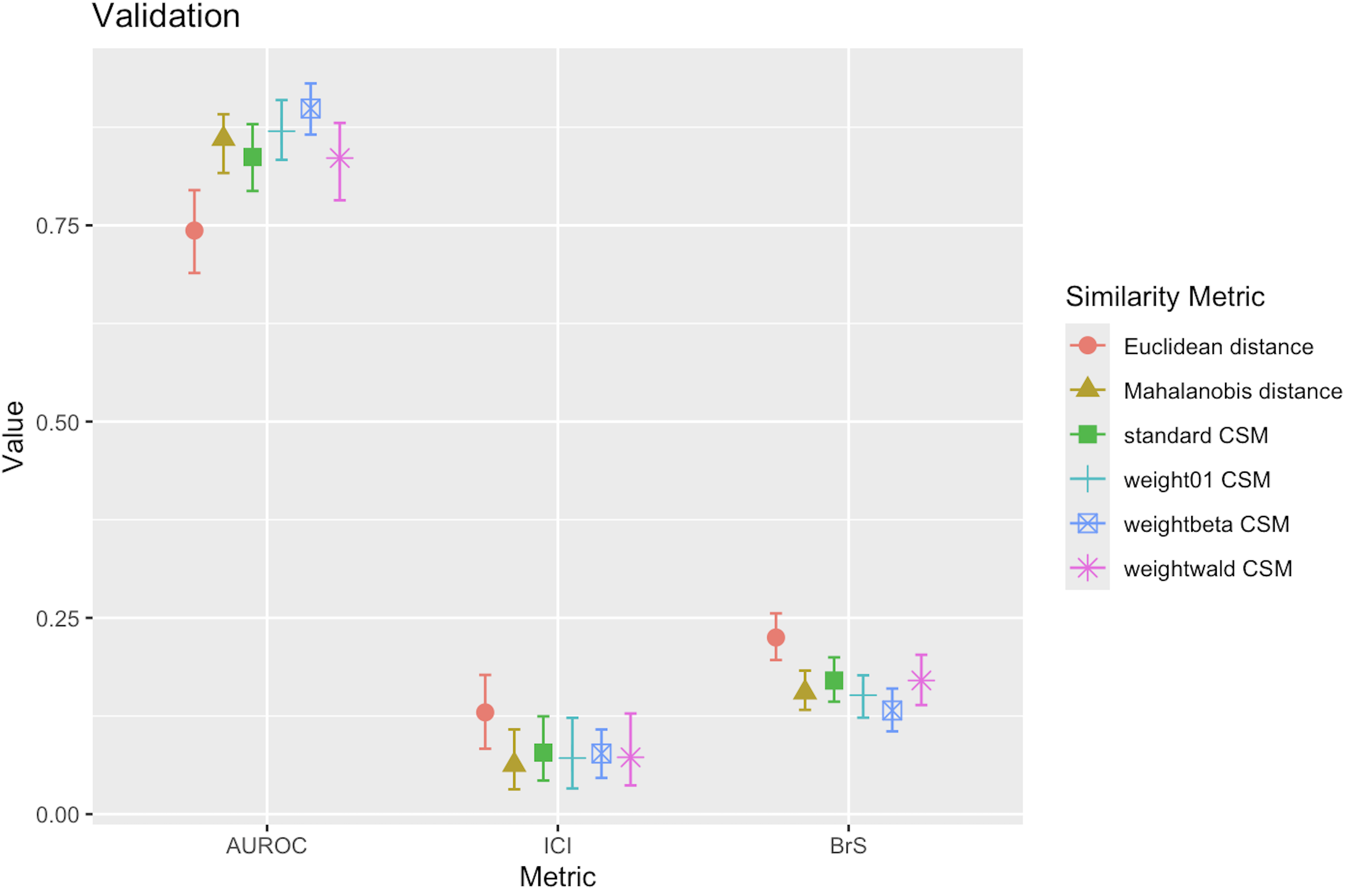}
    \label{fig:plot2}
\end{minipage}

\caption{Performance metrics across different subpopulation proportions for the six similarity measures. The dataset has a reasonably high prevalence with a weak signal (Case 4).}
\label{fig:4plotscase4}
\end{figure}

\begin{figure}[htbp]
\centering

\begin{minipage}[t]{0.49\textwidth}
    \centering
    \includegraphics[width=\textwidth]{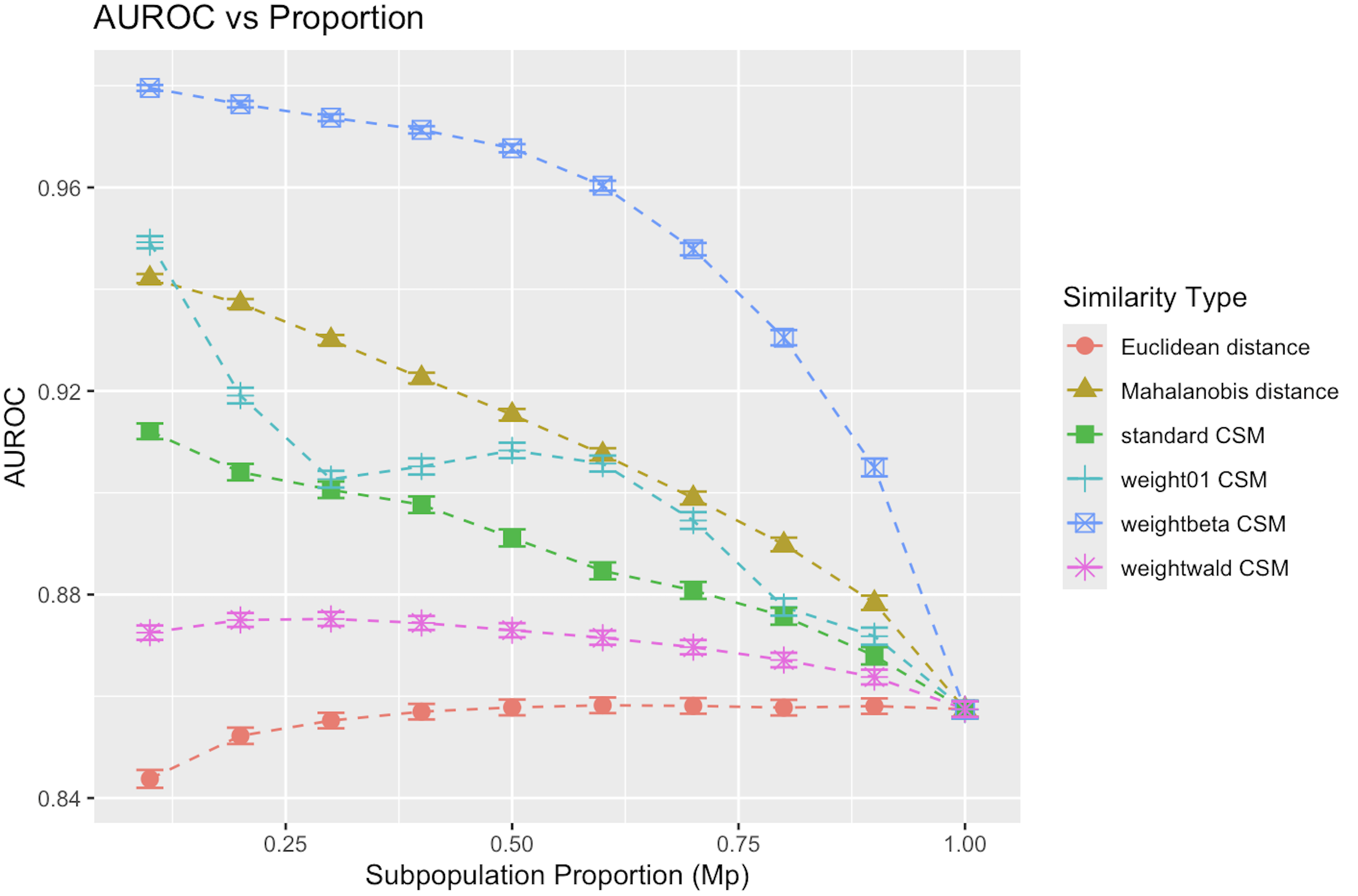}
    \label{fig:plot1}
\end{minipage}
\hfill
\begin{minipage}[t]{0.49\textwidth}
    \centering
    \includegraphics[width=\textwidth]{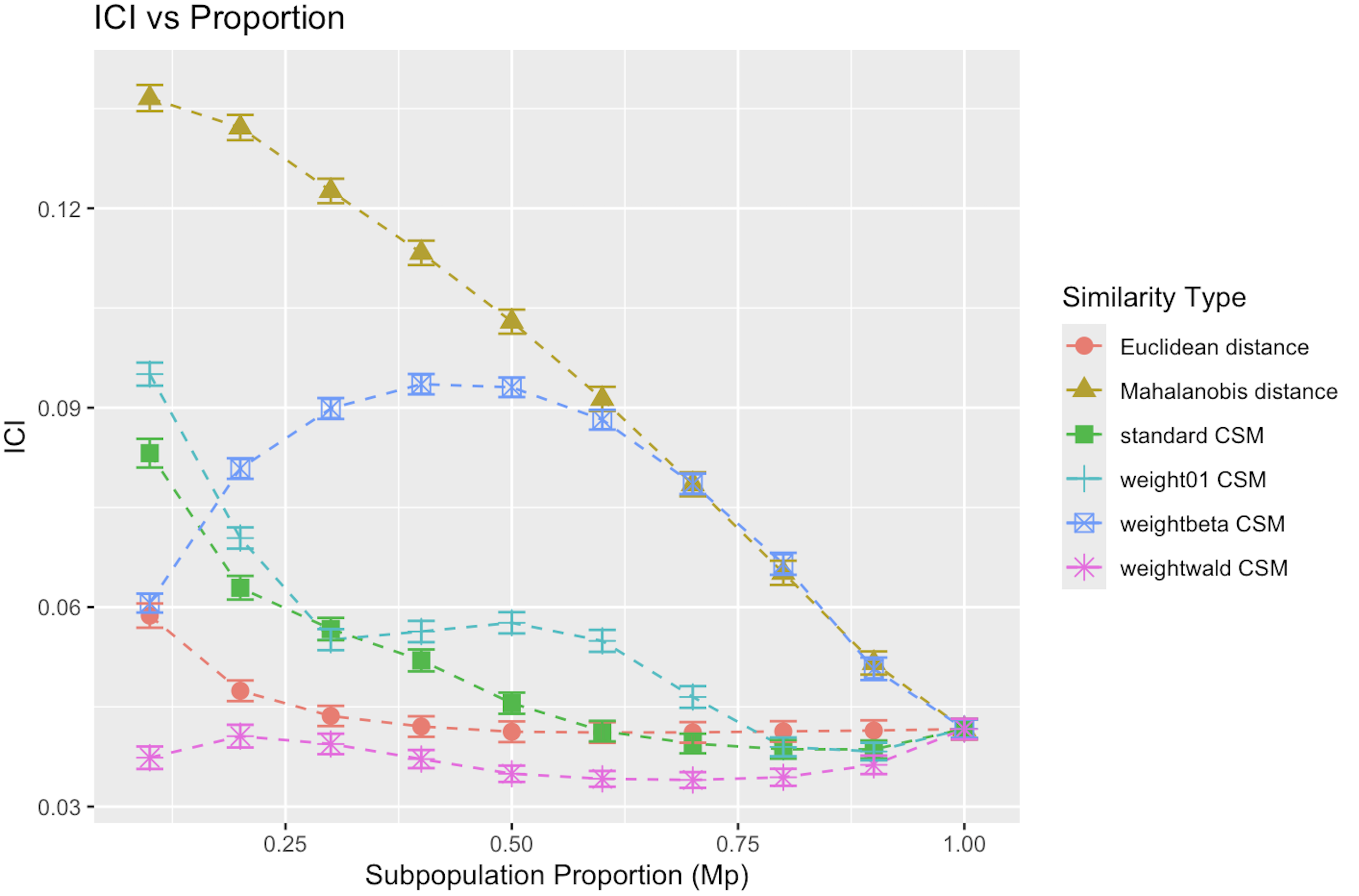}
    \label{fig:plot2}
\end{minipage}

\vspace{0.5cm}

\begin{minipage}[t]{0.49\textwidth}
    \centering
    \includegraphics[width=\textwidth]{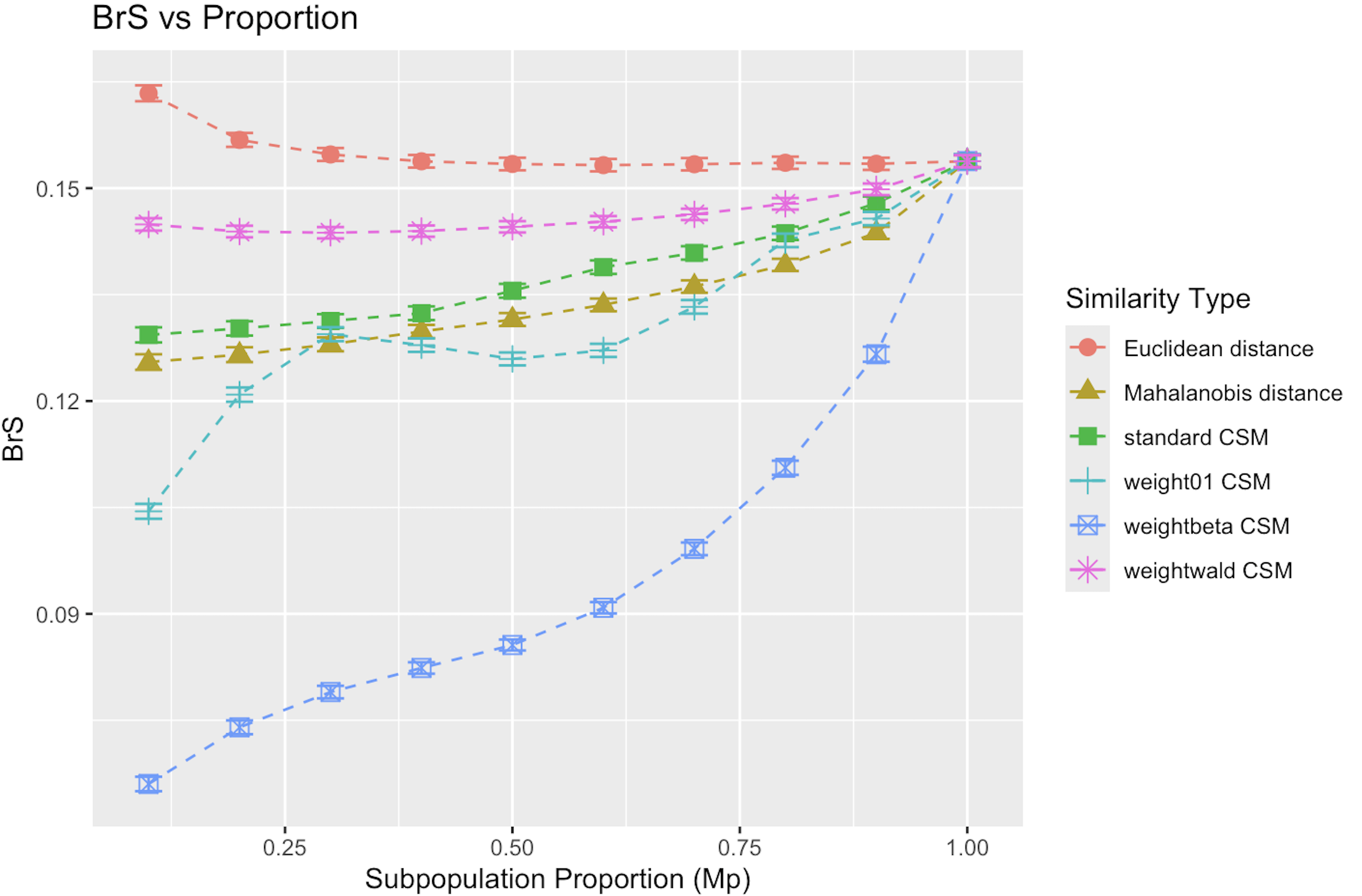}
    \label{fig:plot1}
\end{minipage}
\hfill
\begin{minipage}[t]{0.49\textwidth}
    \centering
    \includegraphics[width=\textwidth]{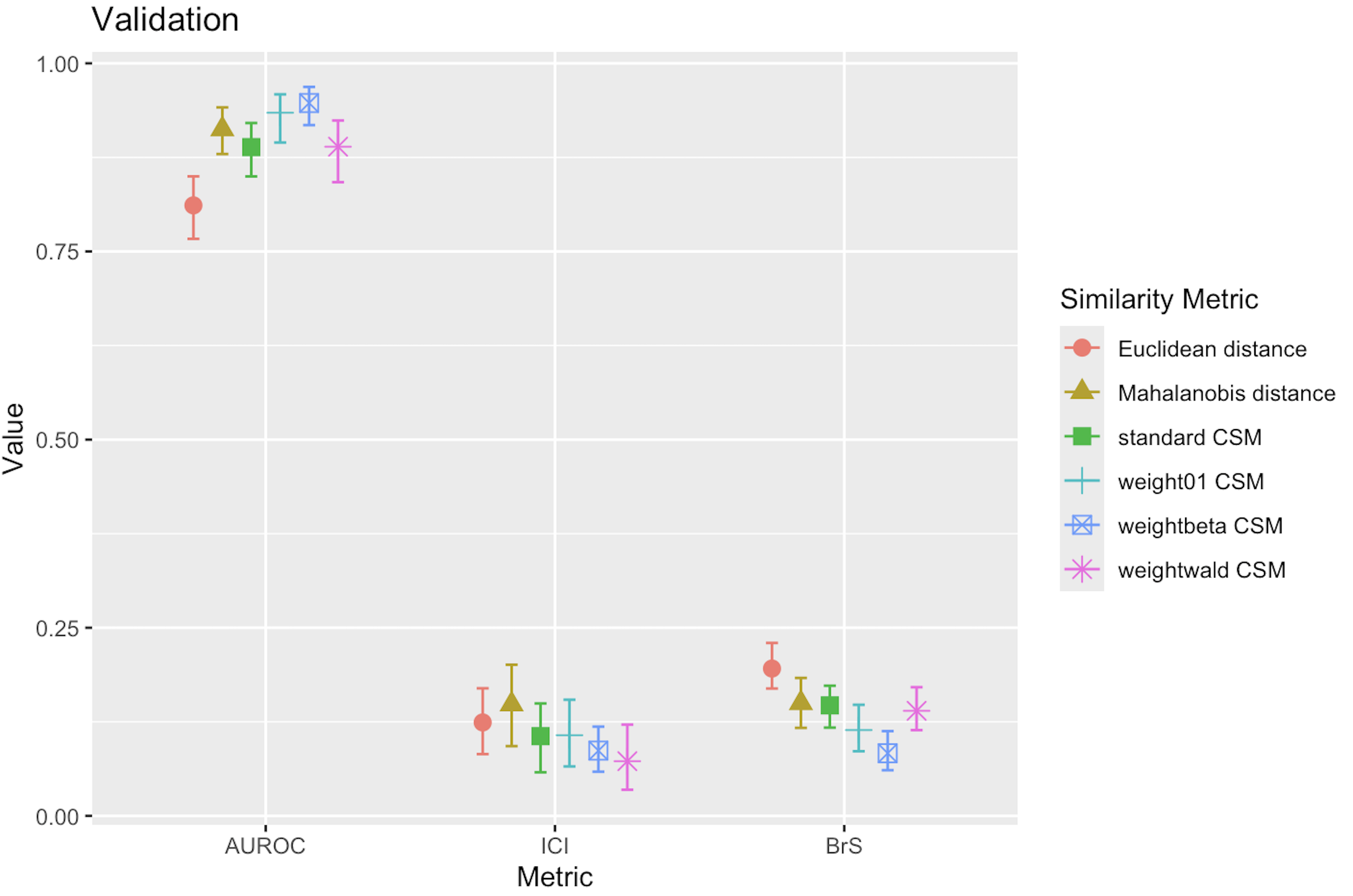}
    \label{fig:plot2}
\end{minipage}

\caption{Performance metrics across different subpopulation proportions for the six similarity measures. The dataset has a reasonably high prevalence with a strong signal (Case 6).}
\label{fig:4plotscase6}
\end{figure}

\end{document}